\documentstyle[tighten,preprint,aps]{revtex}
\newcommand{\sw}{\sin\!\theta_W}
\newcommand{\cw}{\cos\!\theta_W}

\newcommand{\cosb}{\cos\!\beta}
\newcommand{\sinb}{\sin\!\beta}
\newcommand{\sws}{\sin^2\theta_W}

\newcommand{\tws}{\tan^2\theta_W}

\newcommand{\ptau}{P_{\tau}(\tilde{\tau}\rightarrow\tau\tilde{\chi})}

\newcommand{\sti}{\tilde{\tau}_1}
\newcommand \sla[1]{\not{\! {#1}}}
\newcommand{\stii}{\tilde{\tau}_2}
\newcommand{\cost}{\cos\theta_{\tilde{\tau}}}
\newcommand{\sint}{\sin\theta_{\tilde{\tau}}}
\newcommand{\smu}{\tilde{\mu}}
\newcommand{\se}{\tilde{e}}
\newcommand{\dbarchi}{\Delta\bar{\chi}^2}
\newcommand{\stl}{\tilde{\tau}_L}
\newcommand{\str}{\tilde{\tau}_R}
\newcommand{\stau}{\tilde{\tau}}
\newcommand{\mstau}{m_{\tilde{\tau}}}
\newcommand{\msti}{m_{\tilde{\tau}_1}}
\newcommand{\mstii}{m_{\tilde{\tau}_2}}

\newcommand{\mchi}{m_{\tilde{\chi}^0_1}}
\newcommand{\be}{\begin{equation}}
\newcommand{\ee}{\end{equation}}
\newcommand{\een}{\end{mathletters}}
\newcommand{\ben}{\begin{mathletters}}
\newcommand{\beq}{\begin{eqnarray}}
\newcommand{\eeq}{\end{eqnarray}}
\newcommand{\mweak}{M_{\rm weak}}
\newcommand{\mgut}{M_{\rm GUT}}
\begin{document}
\draft
\preprint{\vbox{\baselineskip=14pt%
   \rightline{KEK-TH-486}\break 
   \rightline{KEK Preprint 96--41}
}}
\title 
{Confronting the Minimal Supersymmetric Standard Model with the Study
of Scalar Leptons at Future Linear $e^+e^-$ Colliders} \author{ Mihoko
M.  Nojiri, \thanks{E-mail address: nojirim@theory.kek.jp} Keisuke
Fujii, and Toshifumi Tsukamoto }
\address
{National Laboratory for High Energy Physics (KEK)\\
 Oho 1-1, Tsukuba, Ibaraki
305, Japan 
}
\date{\today}
\maketitle
\begin{abstract} 
Sleptons can easily be found at future linear $e^+e^-$ colliders if
kinematically accessible. Measurements of their masses and decay
distributions would then determine MSSM parameters. This paper
presents a detailed MC study of the production and decay of the
lighter scalar tau lepton, $\tilde{\tau}_1$.  We found that
$m_{\tilde{\tau}_1}$ and $\theta_{\tilde{\tau}}$ (the left-right
mixing angle of $\tilde{\tau}$) would be measured within an error of a
few percent. It is also found that $\tan\beta$ is determinable in some
region of the parameter space through simultaneous studies of
$\tilde{\tau}_1$- and $\tilde{e}$-pair production: the polarization
measurement of the $\tau$ leptons from $\sti$ decays and the $M_1$,
$m_{\tilde{\chi}^0_1}$ determination using $\tilde{e}_R$ pair
production and decay.  We also point out the possibility to determine
$g_{\tilde{B}\tilde{e}_R e}$ through the measurement of the angular
distribution of the $\tilde{e}_R$-pair production. The error on the
coupling is expected to be comparable to its typical SUSY radiative
correction, which is proportional to
$\log(m_{\tilde{q}}/m_{\tilde{l}})$. The radiative correction 
affects $M_1$ and $\tan\beta$ determination, necessitating the 
full 1-loop radiative correction to the $\se_R$ production 
processes.
The implication of these measurements of the MSSM
parameters on selecting models of the origin of supersymmetry breaking
is also discussed.
\end{abstract}
\pacs{PACS number(s): 14.80.Ly, 12.60.Jv, 13.10+q, 13.88.+e}
\noindent
\section{1. Introduction}
The Minimal Supersymmetric Standard Model (MSSM) \cite{SUSY} is one of
the most promising extensions of the Standard Model (SM). It predicts
the existence of superpartners of SM particles (sparticles) below a
few TeV to remove the quadratic divergence which appears in radiative
corrections to the SM Higgs sector.  The model is thus free from the
so--called hierarchy problem inherent in any non-SUSY GUT models. It
should also be noted that the gauge couplings unify very precisely at
high energy in the MSSM, consistent with a SUSY SU(5) GUT
prediction\cite{UNI}.

Supersymmetry is, however, not an exact symmetry of Nature; instead
it should be somehow broken to give a mass difference between each
particle and its superpartner. Various attempts have been made at
explaining the existence of soft SUSY breaking\cite{SUGRA,DNNS}.
Those different models of SUSY breaking lead to different relations
among the soft breaking mass parameters at some high energy scale
$M_{SB}$; this scale could be as high as $M_{pl}$, or as low as
${\cal O}(10^4)$ GeV, depending on the models. Evolving the mass
parameters with the renormalization group equation(RGE) of the model
from $M_{SB}$ to the weak scale $\mweak$, one thus ends up with
different sparticle mass spectra.
 
Precise measurements of masses and interactions of superparticles will
be one of the most important physics targets once they are discovered. 
If the precision reaches a certain level, we will be able to test if a
new particle satisfies relations predicted by supersymmetry. It
will also enable us to measure SUSY breaking mass parameters, and to
discriminate between models of even higher energy scale responsible for
SUSY breaking.

This is indeed the case at proposed future Linear $e^+e^-$
Colliders(LC) operating at $\sqrt{s}=500$ GeV\cite{JLC1,TSUKA,FLC,APPI}, 
which are designed to provide a luminosity in excess of ${\cal L}=30 $fb$^{-1}
$/year\cite{DESY,JLC1,APPI}.  It should also be stressed that the
background from $W$ boson production to SUSY processes can be
suppressed drastically thanks to the highly polarized electron beam
available only at linear $e^+e^-$ colliders\cite{JLC1,TSUKA}.

The production and decay of the lighter chargino, $\tilde{\chi}^-_1$
and scalar leptons, $\tilde{e}$ and $\tilde{\mu}$ at an LC are studied
extensively in previous works\cite{DESY,JLC1,TSUKA,FLC,APPI}. In
particular, it has been shown by Monte Carlo(MC) simulations that some
relations among soft SUSY breaking parameters, which are predicted in
Minimal Supergravity models (MSUGRA) can be tested very
stringently\cite{JLC1,TSUKA,APPI}. It has also been pointed out that one
can verify some SUSY relations, such as that between the off-diagonal
elements of the chargino mass matrix and the mass of the $W$ boson, or
that between gauge boson-fermion-fermion and gaugino-sfermion-fermion
couplings \cite{FLC}.

In this paper, we discuss production and decay of scalar leptons
$\tilde{\tau}$ and $\tilde{e}_R$, and show how various 
MSSM parameters can be measured from  their production only.

The $\tilde{\tau}$ is a very interesting object to study, since its 
mass parameters depend very sensitively on physics at the GUT scale
($\mgut$)\cite{BH}. In SUSY-GUT models the $\tilde{\tau}$ is in the 
same multiplet with $\tilde{t}$ above $\mgut$. Therefore the 
$\tilde{\tau}$ is expected to have a very large coupling, proportional
to the top Yukawa coupling, 
to colour triplet higgs bosons predicted in GUT models. 
Even though all sfermions have equal mass at $M_{pl}$ in MSUGRA models, the 
large Yukawa coupling reduces the $\tilde{\tau}$ mass at $\mgut$ 
compared to those of the other scalar leptons, which might be 
regarded as a signature of quark-lepton unification at the GUT scale.  
This observation implies that the stau can be found earlier than the 
other charged sleptons, which is also phenomenologically interesting.

In order to obtain the GUT scale mass parameters, one has to evolve
the mass parameters at $\mweak$ toward $\mgut$. This requires
knowledge not only of the $\tilde{\tau}$ mass matrix at $\mweak$, but
also of the weak scale $\tau$ Yukawa coupling $Y_{\tau}= -
gm_{\tau}/(\sqrt{2}m_W \cos\beta)$, which is determined by the ratio
of vacuum expectation values $\tan\beta$. $Y_{\tau}$ may have very big
effects on the RG running of $\msti$ for $\tan\beta\sim 50 $: such a
large value of $\tan\beta$ is expected in a minimal $SO(10)$ GUT model
\cite{SOTEN}.

The measurement of $Y_{\tau}$ is known to be difficult\cite{JLC1}, but
it has been pointed out\cite{NOJIRI,NOJIRI2} that the decay
distribution of $\tilde{\tau}$ contains some information on this
coupling.
 
$\tilde{\tau}$ production and decay is different from $\se$ or $\smu$,
because of the non-negligible $\tau$ Yukawa coupling involved in its
mass matrix and interactions. Due to this coupling, $\str$ and $\stl$
mix, and the mass eigenstates are not necessarily current
eigenstates. The same Yukawa coupling appears as a non--negligible
$\tau \tilde{\tau} \tilde{H_1^0}$ coupling, where $\tilde{H_1^0}$ is a
neutral higgsino. This interaction is involved in $\tilde{\tau}$ decay
into a neutralino ($\tilde{\chi}^0_i$) and $\tau$, or a
chargino($\tilde{\chi}^-_i$) and $\nu_{\tau}$, since the
$\tilde{\chi}$'s are mixtures of higgsinos and gauginos. 

Another
feature of $\tilde{\tau}$ decay that distinguishes it from other
slepton decays is that the daughter $\tau$ lepton from the decay
$\tilde{\tau}\rightarrow\tilde{\chi}^0_i\tau$ further decays in the
detector, which enables us to measure the average polarization of the
$\tau$ ($P_{\tau}(\tilde{\tau}\rightarrow\tau\tilde{\chi})$).
The $\tau$ lepton from $\sti$ decay is naturally polarized. The
polarization $P_{\tau}$ in the decay $\tilde{\tau_1} \rightarrow
\tilde{\chi}_i^0\tau$ depends on $Y_{\tau}$.  This dependence arises because
the interaction of gauginos with (s)fermions preserves chirality and
is proportional to a gauge coupling, while the interaction of
higgsinos flips chirality and is proportional to $Y_{\tau}$.
$P_{\tau}$ from decaying $\sti$ reflects the ratio of the chirality
flipping and conserving interactions and is therefore sensitive to
$Y_{\tau}$.

$P_{\tau}$ also depends on the $\stau$ left-right mixing angle
$\theta_{\tilde{\tau}}$, and on the neutralino mixing $N_{ij}$ which
in turn depends on $(M_1, M_2,\mu,\tan\beta)$. $\theta_{\tilde{\tau}}$
can be determined independently from a measurement of the
$\tilde{\tau}$ pair production cross section.
On the other hand, information on $N_{ij}$ must be obtained elsewhere,
for example from $\tilde{e}_R$ pair production and decay. Selectron
pair production involves $t$-channel exchange of neutralinos. By
studying $\tilde{e}_R$ pair production followed by the decay
$\tilde{e}_R\rightarrow e \tilde{\chi}^0_1$, one can thus not only
measure the mass of the $\tilde{\chi}^0_1$ ($m_{\tilde{\chi}^0_1}$),
but also very strongly constrain the gaugino mass parameter
$M_1$. Making use of the measured $P_{\tau}(\tilde{\tau}\rightarrow
\tau\tilde{\chi}^0_1)$ and assuming a GUT relation between $M_1$ and
$M_2$, we can in principle determine all the parameters of the
neutralino mass matrix: $M_1, \mu$, and $\tan\beta$. One purpose of
this paper is to reveal the feasibility of the $\tan\beta$ measurement at
future LC's.

Another aspect of the $\tilde{e}_R$ and $\sti$ measurements is also
treated in this paper. In the high energy limit, $\tilde{e}_R$
production involves $s$-channel exchange of the $U(1)_Y$ gauge boson
$B$ and $t$-channel exchange of its superpartner $\tilde{B}$. The
process turns out to provide clear information on the
$\tilde{B}$-$e_R$-$\tilde{e}_R$ coupling
$g_{\tilde{B}\tilde{e}_Re_R}$. Assuming that the $\tilde{e}_R$ angular
distribution can be reconstructed from that of daughter electrons, we
find that the sensitivity to the coupling
$g_{\tilde{B}e_R\tilde{e}_R}$ would reach ${\cal O}(1\%)$ (which
corresponds to a few \% sensitivity to the production cross
section). The sensitivity is then comparable to the typical radiative
correction to the SUSY relation $g_{\tilde{B}e_R\tilde{e}_R
}=\sqrt{2}g'$, which is proportional to $\log
(m_{\tilde{q}}/m_{\tilde{l}})$. This is the first example where
radiative corrections to couplings involving superpartners might be
measured experimentally.

We also discuss what the
$P_{\tau}(\tilde{\tau}\rightarrow\tau\tilde{\chi}^0_1)$ measurement
implies in the limit where $\tilde{\chi}^0_1$ is dominantly
gaugino. In this limit, the sensitivity to $\tan\beta$ disappears
since no $\tilde{\tau}\tau\tilde{H_1^0}$ interaction is involved in
the $\tilde{\tau}\tau\tilde{\chi}^0_1$ coupling.  However, in this
case, we show that sensitivity to the chiral nature of
$\tilde{\tau}\tau\tilde{B}$ coupling emerges, offering another test of
a supersymmetry relation.

The organization of this paper is as follows. In Sec.2 we review the
physics involved in the $\sti$ mass matrix. In Sec.2.1, we describe
the relation between the weak scale parameters $\msti$, $\mstii$, and
$\theta_{\tilde{\tau}}$ and GUT scale
$\tilde{\tau}$ mass matrices in detail. The importance of measuring
$\theta_{\tilde{\tau}}$ and $\tan\beta$ is stressed there, since it
allows us to check the relation between $m_{\tilde e}$ and
$m_{\tilde{\tau}}$ at $\mgut$. Sec.2.2 is devoted to describing the
procedure to determine $\tan\beta$ from the measurements of $\ptau$
and of $\tilde{e}_R $ pair production. In Sec.2.3, we discuss the
energy distribution of $\tilde{\tau}$ decay products, from which
$\ptau$ and $\msti$ are measured.

Our MC studies of $\tilde{\tau}$ pair production and decay are
described in detail in Sec.3, where one can find our error estimates
on $m_{\tilde{\tau}}$, $\theta_{\tilde{\tau}}$, and $P_{\tau}$ for
$\int {\cal L}dt=100 $fb$^{-1}$. Some preliminary studies have been
given in proceedings reports \cite{NOJIRI2}, where the effects of the
$e^+e^-\tau^+\tau^-$ background were not properly taken into
account. In this paper, we present our final results with an optimized
set of cuts to remove the background, while minimizing acceptance
distortion for parameter fitting.  These cuts are detailed in Sec.3.1.
The results of the fitting are discussed in Sec.3.2.

In Sec.4.1, we define a function called $\Delta\bar\chi^2$, which
allows convenient estimates of errors on MSSM parameters that could be
obtained through fits of $\stau$ and $\se$ decay distributions.
Sec. 4.2 is devoted to the $\tan\beta$ determination from a
simultaneous fit of $\sti$ and $\tilde{e_R}$ production using
$\dbarchi$, demonstrating a unique opportunity to measure $\tan\beta$
if it is large. In Sec.4.3, we go further to determine $\se \ (\stau)$
coupling to neutralinos. Sec.5 then summarizes our results and
concludes this paper.

\section*{2. Physics of $\sti$}
\subsection*{ 2.1 Origin of Supersymmetry 
Breaking and the Mass of $\tilde{\tau}$} $\tilde{\tau}_{L(R)}$ is the
superpartner of $\tau_{L(R)}$, the third generation lepton. This makes
$\tilde{\tau}$ a unique object in the context of SUGRA-GUT
models\cite{BH}.

In minimal supergravity models, SUSY breaking in a hidden sector induces
a universal soft breaking mass $m_0$, a universal gaugino mass $M_0$, and
a universal trilinear coupling $A_0$ through gravitational interactions
at the Planck scale $M_{pl}$.  If the soft breaking masses remain
universal from $M_{pl}$ through $\mgut$, this boundary condition
results in the universality at the weak scale of sfermion soft
breaking masses within the same representation of the $SU(3)\times
SU(2)\times U(1)_Y$ gauge interactions in the MSSM as long as their
Yukawa interactions are negligible: 
\ben\label{e1}\beq 
m_{\tilde L}^2\vert_{weak}&=&(m_0^{GUT})^2 + 0.5 (M_0^{GUT})^2\\ 
m_{\tilde R}^2\vert_{weak}&=&(m_0^{GUT})^2 + 0.15( M_0^{GUT})^2. 
\eeq\een 
Here $m_{\tilde L(R)}$ is the soft breaking mass of the superpartner of a
left(right)-handed lepton, $m_0^{GUT}$ and $M_0^{GUT}$ are the
universal scalar and gaugino masses at $\mgut$.  The model also
predicts the following relations among gaugino soft breaking mass
parameters: 
\be\label{e2} 
M_1= \frac{5\alpha_1}{3\alpha_G} M_0^{GUT},\
\ \ M_2=\frac{\alpha_2}{\alpha_G} M_0^{GUT},\ \ \
M_3=\frac{\alpha_3}{\alpha_G} M_0^{GUT}, 
\ee 
where $M_1$, $M_2$, and
$M_3$ are the masses of $U(1)_Y$, $SU(2)$, and $SU(3)$ gauginos (called
bino, wino, and gluino, respectively). $\alpha_G$ is the gauge
coupling at $\mgut$ where the $\alpha_i$'s unify.

Eq.(\ref{e1}) does not apply for $\stau$,
due to the possibly large $\tau$ Yukawa coupling allowed in MSSM.
The Yukawa interaction of the third
generation fermions is described in the MSSM by the following
superpotential: 
\be\label{e4} 
W_Y=Y_{\tau} H_1 E^c L+ Y_b H_1 D^c Q +Y_t H_2 U^c Q, 
\ee 
where $H_1$($H_2$) is the higgs doublet with
hypercharge $Y=-1/2 (1/2)$ that gives masses to down(up)-type
fermions after $SU(2)\times U(1)_Y$ symmetry breaking: $\langle
H_1\rangle \equiv (\frac{1}{\sqrt{2}} v_1, 0) $ and $\langle
H_2\rangle \equiv (0,\frac{1}{\sqrt{2}}v_2)$.  Fermion masses are thus
not simply proportional to their Yukawa couplings but depend on
$v_{1,2}$ as well: $Y_{\tau}=-gm_{\tau}/(\sqrt{2}m_W \cos\beta)$ where
$\tan\beta\equiv v_2/ v_1$. In a simple $SO(10)$ model where all the
Yukawa couplings are unified at the GUT scale, $Y_{\tau}\sim 1$ at
$M_{GUT}$ and $\tan\beta$ is predicted to be around
50\cite{SOTEN}. For such a large $\tan\beta$, the contribution of the
$\tau$ Yukawa interaction to the RG evolution of $\mstau$ is
non-negligible. $m_{\tilde{\tau}}$ receives a negative radiative
correction going down from $\mgut$ to $\mweak$, leading to a 
mass reduction compared to $m_{\tilde{e}}$.  Numerical values of the
$\tilde{\tau}$ soft breaking masses at $\mweak$ for a unified Yukawa
coupling at the GUT scale $Y_{GUT}= 1$ can be obtained using
\cite{UGLY} \ben\label{e5}\beq m_{\tilde{L}_{\tau}}^2&=&0.53
(M_0^{GUT})^2 -0.12m_{H}\vert_{GUT}^2+0.77m\vert_{GUT}^2\\
m_{\tilde{R}_{\tau}}^2&=&0.15 (M_0^{GUT})^2 -
0.23m_{H}\vert_{GUT}^2+0.55m\vert_{GUT}^2.  \eeq\een Here
$m\vert_{GUT}$ is the universal soft breaking mass of sfermions, and
$m_H\vert_{GUT}$ is the soft breaking mass of higgs bosons at
$M_{GUT}$, which may be different from $m\vert_{GUT}$ in the $SO(10)$
model.\footnote{ The original formula in Ref.\protect\cite{UGLY} contain
$D$-term contributions, which we have neglected here.}
One can subtract Eq.(\ref{e1}) from Eq.(\ref{e5}), after setting
$m_{H}\vert_{GUT}$ and $m\vert_{GUT}$ equal to $m_0^{GUT}$, to single
out the maximal possible effect of the Yukawa RG running from $\mgut$
to $\mweak$. The effect reduces the coefficient of $(m_0^{\rm GUT})^2$ from 1 to
0.32 for $m_{\tilde{R}_{\tau}}^2$ and to 0.65 for
$m_{\tilde{L}_{\tau}}^2$.

Yet another source of the reduction of $m_{\tilde{\tau}}$ is 
left-right mixing\cite{DN}. The mass matrix of a slepton flavor
$(\tilde{l}_L, \tilde{l}_R )$ can be written as
\begin{equation}\label{e6}
{\cal M}^2=\left(\begin{array}{cc}m_{LL}^2 & m_{LR}^2\\ m_{LR}^2&
m_{RR}^2\end{array}\right) =\left( \begin{array}{cc} m_{\tilde{L}}^2 +
m_{l}^2 + 0.27 D & -m_{l}(A_{l} + \mu \tan\beta)\\ -m_{l}(A_l + \mu
\tan\beta)& m_{\tilde{R}}^2 +m_{l}^2 + 0.23D \end{array}\right).\\
\end{equation}
Here $\mu$ is the higgsino mass parameter, $A_{l}$ is the
coefficient of the soft breaking term proportional to
$\tilde{\l}_R^*$-$\tilde{\l}_L$-$H_1$, and $D\equiv -m_Z^2
\cos(2\beta) $. The left-right mixing element ($m_{LR}^2$) is
negligible for the lighter generations. However, for $\stau$, if
$\tan\beta\sim50$, the suppression from a factor of $m_{\tau}$ is
compensated as long as the diagonal mass parameters are ${\cal
O}(m_W)$. The mixing is also non-negligible if 
$m_{\tilde{L}_{\tau}}, m_{\tilde{R}_{\tau}} \ll \mu$.  Mixing makes the
lighter mass eigenvalue $\msti$ lighter than diagonal mass terms. The
mass eigenstates and eigenvalues are expressed as \ben\label{e7} \beq
\left(\begin{array}{c} \sti\\\stii\end{array}\right)
&=&\left(\begin{array}{cc}\cost &\sint\\ -
\sint&\cost\end{array}\right) \left(\begin{array}{c} \stl\\
\str\end{array}\right),\\ m_{\tilde{\tau}_{1,2}}&=&\frac{1}{2}\left[
m_{LL}^2+m_{RR}^2\mp
\sqrt{(m_{LL}^2-m_{RR}^2)^2+4(m_{LR}^2)^2}\right], \\
\tan\theta_{\stau}&=&\frac{\msti^2-m_{LL}^2}{m_{LR}^2}.  \eeq \een
$\sti$ may hence be lighter than $\tilde{e}$, even in a model with a
common soft breaking sfermion mass at $\mweak$.

We learned that determination of $\tan\beta$ 
characterizing the RG running of $m_{\stau}$ 
from $\mgut$ to $\mweak$, and of the weak scale $\stau$ mass matrix
parametrized by $\msti,\mstii,\theta_{\stau}$, is 
necessary to extract $m_{\tilde{L}_{\tau}}$ and $m_{\tilde{R}_{\tau}}$ 
at $\mgut$. 
The values at $\mgut$ are interesting since they sensitively depend on 
the nature of quark-lepton unification,
as has been emphasized recently in Ref.\cite{BH}. The reason is the
following: In simple grand unified models such as supersymmetric
$SO(10)$ or $SU(5)$ models, the $\tau_{R(L)}$ superfield 
is in the same multiplet
as the top quark superfield above $\mgut$. Thus from $M_{pl}$ to
$M_{GUT}$, the $\tau_{R(L)}$ supermultiplet is subject to the same Yukawa
interaction as the top quark. This reduces
$m_{\tilde{R}_{\tau}}$ (and  $m_{\tilde{L}_{\tau}}$) at $\mgut$ from that
at $M_{pl}$ for the $SU(5)$ ($SO(10)$) GUT model\cite{BH}.  The
reduction is predicted as a function of the top Yukawa coupling $Y_t$,
$m_0$, $M_0$, and $A_0$.  $m_{\tilde{R}_{\tau}}$ and
$m_{\tilde{L}_{\tau}}$ could be as light as zero at $\mgut$ for a
large value of $A_0$, even if $m_0^2\neq 0$.

Phenomenologically the MSUGRA-GUT suggests that $\sti$ can be the
lightest charged SUSY particle, thus to be observed first, or might
even be the only SUSY particle to be accessible at the proposed next
generation linear $e^+e^-$ colliders. However, we should stress that
there exist models which predict totally different soft breaking mass
parameters $m_{{\tilde{L}_{\tau},(\tilde{R}_{\tau})}}$. Dine, Nelson,
Nir and Shirman recently constructed a relatively simple model which
dynamically breaks SUSY at some intermediate scale [$\sim10^{6\sim 7}$
GeV](DNNS model) \cite{DNNS}. The breaking is then transferred to our
sector by a $U(1)_Y$ gauge interaction, whose scale $M_{m}$ is ${\cal
O}(10^4)$ GeV.  Its prediction on the gaugino mass parameters turns
out to be the same as that of the MSUGRA model. This is not the case
for the slepton masses, which are predicted to be common to
($\tilde{l}_L, \tilde{\nu}_l$) and $\tilde{l}_R$, respectively at
$M_{m}$: 
\ben\label{e3}\beq 
\tilde{m}_{\tilde{L}}^2&\propto& \frac{3}{4} 
\left(\frac{\alpha_2}{4\pi}\right)^2
+ \frac{5}{3}\left(\frac{1}{4}\right)^2\left(\frac{\alpha_1}{4\pi}
\right)^2  \label{e3a} \\ 
\tilde{m}_{\tilde{R}}^2 &\propto& \frac{5}{3}
\left(\frac{1}{2}\right)^2 \left(\frac{\alpha_1}{4\pi}\right)^2
\label{e3b}.  
\eeq\een 
Unlike in the SUGRA-GUT model, the slepton masses do not run too much
from $M_m$ to $M_{weak}$, as $M_{m}$ is considerably closer to
$M_{weak}$ and there is no strong Yukawa interaction involved at these
energy scales.\footnote{Even if we take a very large $Y_\tau$ at the
GUT scale, the weak scale value of $Y_{\tau}$ is smaller compared to
that of $Y_{b}$ and $Y_{t}$.}

The determination of $m_{\tilde{L}_{\tau}}$, 
$m_{\tilde{R}_{\tau}}$, and
$m_{\tilde{e}(\tilde{\mu})}$ at the GUT scale  would therfore give us a
good handle to distinguish the MSUGRA and DNNS models, or if the scale
of SUSY breaking is below or above the GUT scale;  it
is not enough to only observe $m_{\tilde{\tau}}<m_{\tilde{e}}$, but
$\tan\beta$ and $\tilde\theta_{\stau}$ must be measured to determine
$m_{\tilde{L}_{\tau}}$ and $m_{\tilde{R}_{\tau}}$ at $\mgut$.  This
shows the importance of precision studies of production and decay of
$\sti$ at future LC's.  We discuss in the next subsections how we can
measure these parameters using $\tilde{\tau}_1$ pair production and
decay.

\subsection*{ 2.2 Determination of 
MSSM Mass Parameters from Production and Decay of Sleptons}

Information on $\theta_{\stau}$ and $\tan\beta$ can be extracted
solely from the production and decay of $\stau$\cite{NOJIRI,NOJIRI2}.
In this subsection we sketch our strategy to do this. The
determination of $\msti$ will be discussed in Sec. 2.3.

A $\stau$ decays into a chargino $\tilde{\chi}^-_i$ (i=1,2) plus a
$\nu_{\tau}$, or a $\tilde{\nu}_\tau$ plus a $W^-$ or a $H^-$, or a
neutralino $\tilde{\chi}^0_i$(i=1,...,4) plus a $\tau$.  Here the
neutralinos are some mixtures of the neutral components of gauginos
and higgsinos ($\tilde{B}$, $\tilde{W}$, $\tilde{H}_1^0$ and
$\tilde{H}_2^0$ ), and the charginos are some mixtures of the charged
components.  Throughout this paper we assume that the lightest SUSY
particle (LSP) is the lightest neutralino $\tilde{\chi}^0_1$.  Due to
R parity conservation in the MSSM the LSP is stable and escapes from
detection.  The decay products of any SUSY particle contain at least
one $\tilde{\chi}^0_1$.

When both of the pair-produced $\sti$ decay into $\tilde{\chi}^0_1+$
$\tau$, the event yields a simple acoplanar $\tau^+\tau^-$ final
state. If $\sti$ decays to heavier $\tilde{\chi}^0_i$ are allowed, the
event might contain associated jets or leptons. Notice that if the
$\tilde{\chi}^0_i$ decays into $\nu\bar{\nu}\tilde{\chi}^0_1$, the
event has the same signature as that of
$\sti\rightarrow\tau\tilde{\chi}^0_1$. If one or both $\sti$'(s) decay
into $\tilde{\chi}^{\pm}_i$ +$\nu_{\tau}$ the event results in only
one or zero $\tau$ lepton + jets or a lepton + missing momentum.

The $\sti$-to-ino decay branching ratios depend on the scalar tau
mixing $\theta_{\stau}$ and the parameters of the -ino sector ($M_1$,
$(M_2)$, $\mu$, $\tan\beta$). The measurement of the $\sti$ branching
ratios might give us extra information on these parameters, but the
existence of various decay modes also makes the analysis of $\sti$
production very complicated. This point has been discussed in previous
works\cite{NOJIRI} in detail, and we will not repeat it here.

Hereafter we concentrate on the case in which the $\stau_R$ is the
second lightest SUSY particle and decays exclusively into
$\tau\tilde{\chi}^0_1$.  Fig.1 shows the interactions of the neutral
components of gauginos and higgsinos with $\stau_R$ and $\tau$. The
interaction is completely fixed by supersymmetry.  Namely, the
coupling of the $\stau_R$ to $\tilde{B}$ is proportional to the U(1)
gauge coupling $g_{1}$, while the coupling to $\tilde{H}_1^0$ is
proportional to $Y_{\tau}$. The two interactions have different
chirality structure.  The (super-)gauge interaction is chirality
conserving, while the (super-)Yukawa interaction flips chirality (In
the figure, the arrows next to the $\tilde{\tau}$ and $\tau$ lines
show flow of chirality). Since the polarization of the $\tau$ lepton
$P_{\tau}(\tilde{\tau}_R \rightarrow \tau\tilde{\chi}^0_1)$ measures
the ratio of the chirality flipping and the conserving interactions,
it is sensitive to $\tan\beta$.

As we mentioned already, the gauginos and higgsinos are not mass
eigenstates, but they mix to form the neutralino mass eigenstates
$\tilde{\chi}^0_i(i=1,.., 4)$. The $\str$ and $\stl$ also mix. Hence
the $\tilde{\chi}^0_i\sti\tau$ couplings depend not only on
$\tan\beta$ but also on the stau mixing $\theta_{\stau}$ and the
neutralino mixing $N_{ij}$, where $N_{ij}$ is defined by
$\tilde{\chi}^0_i=N_{i1}\tilde{B}+N_{i2}\tilde{W}+N_{i3}
\tilde{H}_1+N_{i4}\tilde{H}_2$.  Therefore, the measurement of
$P_{\tau}$ alone can not uniquely determine $Y_{\tau}$ unless
$\theta_{\stau}$ and $N_{ij}$ are specified. For example, in the limit
where the lightest neutralino $\tilde{\chi}^0_1$ is a pure bino
state ($N_{11}\rightarrow 1$) and in the
$\tilde{\tau}_1\rightarrow\tilde{\tau}_{R(L)}$ limit, $P_{\tau}$ is
expressed as 
\ben\beq\label{e8}
P_{\tau}\left(\sti\rightarrow\tilde{B}\tau\right)&=&
\frac{4\sin^2\theta_{\stau} - \cos^2\theta_{\stau}}
{4\sin^2\theta_{\stau} + \cos^2\theta_{\stau}} \label{e8a}\\
P_{\tau}\left(\tilde{\tau}_R\rightarrow\tilde{\chi}^0_1\tau\right)&=&
\frac{\left(\sqrt{2}N_{11}\tan\theta_W\right)^2 -
\left(Y_{\tau}N_{13}\right)^2}
{\left(\sqrt{2}N_{11}\tan\theta_W\right)^2
+\left(Y_{\tau}N_{13}\right)^2} \label{e8b}\\
P_{\tau}\left(\tilde{\tau}_L\rightarrow\tilde{\chi}^0_1\tau\right)&=&
\frac{\left(\sqrt{2}Y_{\tau}N_{13}\right)^2-g^2
\left(N_{12}+N_{11}\tan\theta_W\right)^2}
{\left(\sqrt{2}Y_{\tau}N_{13}\right)^2+
g^2\left(N_{12}+N_{11}\tan\theta_W\right)^2}, 
\eeq\een
respectively. In the gaugino dominant limit, $P_{\tau}$ does not
depend on $\tan\beta$ as expected. On the other hand, if $N_{13}$ is
non-negligible, $P_{\tau}$ depends on $\tan\beta$,
but how it depends differs as $\sin\theta_{\stau}$ varies from 0 to
1. The interactions involving $\tilde{\tau}$, $\tilde{\chi}_i^0$
$\tilde{\chi}_i^+$, and the dependence of $P_{\tau}$ on these
interactions are listed in appendix A, together with the definitions
of neutralino and chargino mixing angles.

Now let us turn to the determination of $\theta_{\stau}$ and $N_{ij}$.

\noindent
{\it$\bullet$ $\sti$ Mixing Angle $\theta_{\stau}$}:
 
Since a polarized electron beam will be available at future linear
$e^+e^-$ colliders, the mixing angle $\theta_{\stau}$ can be
determined by the measurement of the production cross section for
$e^+e^-\rightarrow\sti^+\sti^-$\cite{NOJIRI}. This can easily be seen
by taking the limit $m_Z\ll\sqrt{s}$ and $P_e=+1$.  In this limit, the
$\stau$ production solely proceeds through the exchange of the
$U(1)_Y$ gauge boson $B$. The hypercharge for $\tilde{\tau}_{L(R)}$ is
$-1/2$($-1$), thus $\sigma(\tilde{\tau}_R)$ $\sim 4
\sigma(\tilde{\tau}_L)$. Though the cross section also depends upon
$\msti$, it can be separately extracted from the energy distribution
of $\stau$ decay products or from a threshold scan\cite{NOJIRI}. 

\noindent
{\it $\bullet$ Neutralino Mixing Angles $N_{ij}$}\cite{TSUKA,NOJIRI2}:

The neutralino mixing $N_{ij}$ depends on $M_1, M_2,\mu$, and
$\tan\beta$. If we assume the GUT relation
$M_1=5\alpha_1/(3\alpha_2)\cdot M_2$, we can determine two out of the
three parameters using $\tilde{e}^+_{R}\tilde{e}^-_R$ pair production
as we will discuss below.\footnote{For any numerical calculation in
this paper we assume the GUT relation, though the ratio might be
determined model independently, using chargino
production\protect\cite{TSUKA,FLC}.} Combining it with the
$\sigma_{\stau\stau}$ and $P_{\tau}$ measurements, one can then
determine all the parameters of the neutralino mass matrix in
principle.

The $\tilde{e}_R$ pair production proceeds though the $s$-channel
exchange of gauge bosons and the $t$-channel exchange of
neutralinos. We list the amplitudes for the $e^+e^-\rightarrow
\tilde{e}^+_R\tilde{e}^-_R$ production in Appendix B. In the limit
$M_1,\mu\gg m_Z$, $\sqrt{s}\gg m_Z$ and $P_e=+1$, the amplitude
reduces to \be\label{e9} i{\cal M}\rightarrow i\beta_f
g^2\tan^2\theta_W\sin\theta\left[1-
\frac{4}{1-2\cos\theta\beta_f+\beta_f^2+4M_1^2/s}\right], \ee where
$\theta$ and $\beta_f$ are the polar angle and the velocity of the
$\tilde{e}_R^-$. The first term in the square bracket corresponds to
the $s$-channel exchange of gauge bosons, and the second term to the
$t$-channel $\tilde{B}$ exchange. One can see that only the
interaction with the $U(1)_Y$ gauge boson ($B$) and the gaugino
($\tilde{B}$) is relevant in the limit, since $\tilde{e}_R$ is an
SU(2) singlet.
 
The differential cross section
$d\sigma(e^+e^-\rightarrow\tilde{e}_R^+\tilde{e}_R^-)/\cos\theta$ is
very sensitive to $M_1$. In Ref.\cite{TSUKA},
$\tilde{e}^+_R\tilde{e}^-_R$ production and their subsequent decays
$\tilde{e}^{\pm}_R\rightarrow e^{\pm}\tilde{\chi}^0_1$ have been
studied in detail. It was pointed out that the three-momentum of
$\tilde{e}_R$ can be derived from the momenta of the final-state
electron pair with a twofold ambiguity, provided that
$m_{\tilde{e}_R}$ and $m_{\tilde{\chi}^0_1}$ are known. The $\se_R$
and $\tilde{\chi}^0_1$ masses can, on the other hand, be determined
from the energy distribution of the electrons with an error of
{\cal O}(1GeV). The study of $\tilde{e}_R$ therefore provides two out of the
three parameters of the neutralino sector.
The remaining freedom of $\tan\beta$ can then be fixed by $P_{\tau}$
$(\sti\rightarrow \tau)$.

\vskip 1cm

In order to illustrate how the above procedure works, we calculate
the cross section contours for $e^+e^-\rightarrow
\tilde{e}^+_R\tilde{e}^-_R$, and
$P_{\tau}(\tilde{\tau}_R\rightarrow\tau\tilde{\chi}^0_1)$, fixing
$m_{\tilde{\chi}^0_1}$ at 100 GeV and varying $M_1$ and $\tan\beta$.
Curves in the $M_1$-$\mu$ plane which satisfy the $\tilde{\chi}^0_1$
mass constraint are shown in Fig.2 for different values of
$\tan\beta$.  With the mass constraint, one can specify the position
in the parameter space of the neutralino sector by $M_1$ and
$\tan\beta$, up to a twofold ambiguity of positive and negative $\mu$
solutions for the large $M_1$ and $\tan\beta$ region, or up to a
threefold ambiguity of two solutions in the negative $\mu$ and one
solution in the positive $\mu$ regions for small values of $M_1$ and
$\tan\beta$.\footnote{The ambiguities in $\mu$ might be removed for
$\tan\beta<10$ by measuring other processes such as chargino
production and decay.}

The $\sigma_{\tilde{e}_R^+\tilde{e}_R^-}$ contours corresponding to
positive and negative $\mu$ solutions are shown in Fig.3a) as the
solid and dotted lines, respectively. The difference of the two
solutions is bigger for smaller $\tan\beta$. For $\tan\beta>10$, the
difference becomes negligible.

The dependence on $\tan\beta$ is also mild for
$\sigma(e^+e^-\rightarrow \tilde{e}_R^+\tilde{e}_R^-)$, since it only
comes in though the effect of the gaugino-higgsino mixing, which is
suppressed by $m_Z^2 / \max(M_1,\mu)^2$ and $\sin^2\theta_W$, compared
to the leading term. The effect is visible for $\tan\beta<5$ but it
essentially vanishes for $\tan\beta>10$. On the other hand, the cross
section is very sensitive to $M_1$ as expected: it decreases
monotonically with increasing $M_1$, and turns out to be extremely
small when $M_1\sim \sqrt{s}$, where the $t$-channel and $s$-channel
diagrams almost cancel each other.

Fig.3b) is a contour plot of $P_{\tau}$ ($\str\rightarrow
\tilde{\chi}^0_1\tau$) in the $M_1$-$\tan\beta$ plane.  As long as
$M_1$ is not very close to $m_{\tilde{\chi}^0_1}$, the polarization
depends on $\tan\beta$ sensitively in the region of the parameter
space shown in the figure. As one can easily see from Figs.3a) and
3b), if we know $M_1$ precisely from the $\tilde{e}_R$ production
cross section, we can extract $\tan\beta$ by measuring
$P_{\tau}$$(\str\rightarrow \tilde{\chi}^0_1\tau)$ unless
$M_1\sim\mchi$. Notice when $M_1\sim\mchi$, the lightest
neutralino is gaugino dominant and there is no significant Yukawa
coupling involved in $P_{\tau}$ as shown in Eq.(\ref{e8a}). Therefore
we cannot expect any sensitivity to the tau Yukawa coupling in such a
region of parameter space. While for $M_1\gg\mchi$, the lightest
neutralino is higgsino dominant and $\tilde{\chi}^0_1$ has significant
higgsino component. In such case, some sensitivity to $\tan\beta$ is
expected for a moderate value of $\tan\beta$ where the first and the
second terms of the numerator of Eq.(\ref{e8b}) are comparable.

Notice that in Fig.3 we did not exclude the region forbidden by the
Minimal Supergravity Model. In MSUGRA, one has to require the square
of any scalar mass parameter be positive at $M_{pl}$ to prevent the
potential from being unbounded from below there.  This condition leads
to the following inequalities at $\mweak$:
$m_{\tilde{e}_{R}}\geq\sqrt{0.87M_1^2 +0.23D}$ (See Eq.(\ref{e1})). For
instance $m_{\tilde{e}_R}=200$ GeV requires $M_1<215$ GeV.  If we find
$M_1>215$ GeV, it will immediately bring us to conclude that the SUSY
breaking scale $M_{SB}$ is much lower than $M_{GUT}$, and above
$M_{SB}$ the theory is different from the MSSM. In the following
numerical calculations, we will not assume the positivity of the
scalar potential at $\mgut$, since the existence of models with
$M_{SB}\ll M_{GUT}$ is not excluded.  The DNNS model is an example of
such a model with $M_{SB}\ll M_{GUT}$, although their resulting
slepton and gaugino masses at the low energy scale are consistent with
the positive scalar mass requirement at $M_{GUT}$.\footnote{Another 
important sets of constraints could be obtained from
requiring the scalar potential neither be  unbounded from 
below(UFB) nor have charge or 
color breaking(CCB) minima  deeper than the
standard minimum \protect\cite{CASAS}. One would then find 
strong constraints 
on $m_{\tilde{\tau}_L}$ and $m_{\tilde{\tau}_R}$ at the weak scale,
depending on $\mu$ and $m^2_{H_2}$. In this paper, we do not consider
these constraints, since  we will not specify $m_{\tilde{\tau}_2}$,
$m^2_{H_2}$, and $\mu$ in the later analysis.}

\subsection*{ 2.3 Energy Distribution of $\tilde{\tau}$ Decay Products} 

In this subsection, we discuss the measurements of
$\msti$, $\sigma_{\stau\stau}$, and $P_{\tau}$ from the decay
distribution of the $\tau$ leptons from the $\sti$ decays. As discussed in
Sections 2.1 and 2.2, these parameters are important to determine both 
the $\stau$ mass matrix and $\tan\beta$. In Ref.\cite{NOJIRI}, one of 
us (MMN) has proposed measurements using the pion energy distribution in 
$\tilde{\tau}\rightarrow \tilde {\chi}^0_1\tau\rightarrow
\tilde{\chi}^0_1\pi\nu_{\tau}$.  In this paper, 
we discuss measurements using the decay chain $\tilde{\tau}
\rightarrow\tilde{\chi}^0_1\tau\rightarrow\tilde{\chi}^0_1\rho\nu_\tau$, 
since we find  this channel advantageous over 
$\tau\rightarrow \tilde{\chi}^0_1\pi\nu_{\tau}$, as explained below.

First consider the primary decay $\tilde\tau\to\tau\tilde\chi_1^0$.   
The kinematics is analogous to the $\tilde e$ or $\tilde\mu$ cases 
studied in Ref.~\cite{TSUKA}.  The $\tau$ energy distribution is 
flat between the endpoints given by
\be\label{e10}
E^\tau_{\rm max(min)}=\frac{E_\tau^{*} \pm p_\tau^{*} 
\beta_{\tilde\tau}}{\sqrt{1-\beta^2_{\tilde\tau}}}, \ee 
where $E_\tau^{*}$ and $p_\tau^{*}$ are
the $\tau$ energy and momentum in the parent $\tilde\tau$ rest frame:
\be E_\tau^{*} = \frac{m_{\tilde\tau}^2- m_{\tilde\chi^0_1}^2 
+ m_\tau^2}{2m_{\tilde\tau}},\qquad 
p_\tau^{*} = \sqrt{( E_\tau^{*})^2 - m_\tau^2},  \ee
and $\beta_{\tilde\tau}=(1-4m_{\tilde\tau}^2/s)^{1/2}$ is the
$\tilde\tau$ velocity in the laboratory frame.  Knowledge of the two
endpoint energies allows us to determine $m_{\tilde\tau}$ and
$m_{\tilde{\chi}^0_1}$, unless $\beta_{\tilde{\tau}}$ is very
close to one.

However, the $\tau$ decays into $A \nu_{\tau}$, where 
$A= e\nu_e, \mu\nu_{\mu}, \pi, \rho, a_1$, {\it etc.}, and 
$\rho^{\pm}$ further decays into 
$\pi^{\pm}\pi^0$ and $a_1^{\pm}$ to $\pi^{\pm}\pi^{\pm}\pi^{\mp}$
or $\pi^{\pm}\pi^0\pi^0$.  Thus the signature of the $\stau^+\stau^-$
production is an acoplanar two-jet event with low multiplicity.

In the limit $E_{\tau}\gg m_{\tau}$, the decay products keep the original 
$\tau$ direction.  However, the visible energy is smaller 
since some of the $\tau$'s energy is carried away by neutrinos.  
In order to determine $m_{\tilde{\tau}}$ and
$m_{\tilde{\chi}^0_1}$, one must reconstruct the original $\tau$ 
endpoint energies from the energy distribution of the decay products.  
In Figs.~4a) and b), we show the energy distributions of the $\rho$ and 
$\pi$ from a decaying $\tau_{R(L)}$ with a fixed $E_\tau$ in the limit 
$E_{\tau}\gg m_{\tau}$ \cite{HAGI}.  
The energy distributions in the c.m.\ frame is 
obtained by convoluting these distributions with the $\tau$ energy 
distribution, which we show in Figs.~4c) (for $\stau_1\to\tau^-_R$) 
and 4d) ($\stau_1\to\tau^-_L$)
for a representative set of parameters $\msti=150$ GeV, 
$m_{\tilde{\chi}}=100$ GeV, and $\sqrt{s}=500$ GeV.  


The $\pi$ energy distribution depends on $P_\tau$ very strongly.  
It is harder (softer) for a $\pi^-$ from a $\tau^-_{R(L)}$ (see 
Figs.~4c) and d)), due to angular momentum conservation. 
However,  a substantial correlation between $P_{\tau}$ 
and the number of identified events is expected due to the inevitable 
$E_{vis}$ and $\sla{P}_T$ cuts to remove the $e^+e^-\tau^+\tau^-$ 
background \cite{DESY}.  As we will see in Sec.~3, applying these cuts 
drastically reduces events in lower energy region 
$E\lesssim E_{\rm min}^\tau$, where most of the events reside 
for $P_{\tau}\sim -1$. (The maximum and minmum energies of the  
original $\tau$ lepton $E^{\tau}_{\rm max(min)}$ are shown in Figs. 
4c) and d).)  
If $E^{\tau}_{\rm min}\lesssim \not{\! {P}}^{\rm cut}_T$, 
it is thus hard to measure 
the energy distribution precisely, which results in large errors on 
$E^{\tau}_{\rm min}$ and $P_{\tau}$. This uncertainty on $P_{\tau}$ also
affects the determination of $E^{\tau}_{\rm max}$, as the energy 
distribution near $E^{\tau}_{\rm max}$ depends on $P_{\tau}$ strongly.  
Finally, the acceptance depends on $P_{\tau}$, giving extra uncertainty 
on the $\sti$ total cross section measurement.

The $\rho$ mode is preferable to the $\pi$ mode in these aspects.  
The dependence of the energy distribution of
$\rho$ mesons on $P_{\tau}$ is mild, since kinematics forbids low
energy $\rho$ mesons.  The energy distribution is peaked near
$E^{\tau}_{min}$ for any $P_{\tau}$. The $P_\tau$ dependence of 
the energy distribution near $E^{\tau}_{max}$ is also 
moderate.  Because of this pseudo-$P_{\tau}$-independence, we can carry
out the determinations of $m_{\tilde{\tau}}$, 
($m_{\tilde{\chi}^0_1}$), and the cross section without any strong 
correlation to $P_{\tau}$.

Furthermore, the polarization of the $\rho$ meson depends on $P_\tau$
very strongly, which can be seen in the distributions of $\rho_{L(T)}$
in Fig.~4c/d) (dashed lines). Namely, a $\tau_R^-$ decays mostly to a
longitudinally polarized $\rho$ meson ($\rho_L$) and a $\tau_L^-$
decays mostly to a transversally polarized $\rho$ meson
($\rho_T$). One can thus determine $P_{\tau}$ by measuring $P_{\rho}$,
which in turn can be determined from the distribution of the $\rho$
decay products. A $\rho^{\pm}$ decays into $\pi^{\pm} \pi^0$, and the
distribution of $E_{\pi^{\pm}}$ in the $\rho_{L(T)}\rightarrow$
$\pi^{\pm}\pi^0$ decay is a very simple function of $z_c\equiv
E_{\pi^{\pm}}/E_{\rho}$, where $E_{\rho}$ is the total energy of the
jet to which the $\pi^{\pm}$ belongs, and can be written in the
following form\cite{HAGI}:
\ben\label{e11}\beq
d\Gamma(\rho_T\rightarrow 2\pi)/dz_c&\sim&
2z_c(1-z_c)-2m_{\pi}^2/m_{\rho}^2, \\ d\Gamma(\rho_L\rightarrow
2\pi)/dz_c &\sim& (2z_c-1)^2, \eeq\een 
where we have ignored terms ${\cal O}(m^2_{\rho}/E^2_{\rho})$ but
retained  ${\cal O}(m^2_{\pi}/m^2_{\rho})$ contributions, and $z_c$ is
in the range $(1-\beta_{\pi})/2\leq z_c\leq (1+\beta_{\pi})/2 $, with
$\beta_{\pi}=\sqrt{1-4m_{\pi}^2/m_{\rho}^2}$.

By fitting the $z_c$ distribution together with the $E_{\rm jet}$
distribution, one can determine both $P_{\tau}$ and $\msti$. The
error from the small $P_\tau$ dependence of the $E_{\rm jet}$
distribution is reduced by the simultaneous use of the $z_c$
distribution.

\section*{3. Monte Carlo Simulations}
\subsection*{3.1 Event Selection}
In this section, we investigate the feasibility of the $\stau$ studies
outlined above at future linear $e^+e^-$ colliders.

As discussed in the previous section, measuring the $E_{\rho}$
distribution of the cascade decay $\sti\rightarrow\tau\rightarrow\rho$
and the $z_c \ (\equiv E_{\pi^{\pm}}/E_{\rho})$ distribution of the
subsequent $\rho$ decay $\rho\rightarrow \pi^\pm\pi^0$, one can
determine $\msti$, $P_{\tau}$, and $\theta_{\stau}$. However, in order
to measure these parameters, one has to introduce cuts to control
backgrounds. We also have to reconstruct $\rho$, $a_1$, or $\pi$ from
$\sti$ decays with minimum mis-ID probability among  these
channels. The reconstruction efficiency heavily depends on detector
performance, necessitating Monte Carlo(MC) simulations with realistic
detector and machine parameters. In this subsection we dicuss the
dominant background from $e^+e^-\rightarrow e^+e^-\tau^+\tau^-$ 
 and present our
cuts and detector set up to reduce it. We also define  our cuts to
identify $\rho$ and $a_1$ and MC-examine the contamination
due to misidentifications. The result of the fit to MC data
after the cuts will be presented in Sec 3.2.

In the following we study a sample case of $\stau_1^+\stau_1^-$ pair
production followed by exclusive $\sti^{\pm}\rightarrow 
\tau \tilde{\chi}^0_1$ decays. We
will not treat the other decay processes:
$\tilde{\tau}^{\pm}\rightarrow $ $\tau^{\pm}\tilde{\chi}^0_i,(i\ge 2)$
or $\nu_{\tau}\tilde{\chi}^\pm_i$ where the expected event signatures
are much more complicated. The helicity amplitudes for
$\stau_1^+\stau_1^-$ production and their subsequent decays into
$\tau\tilde{\chi}^0_1$ are calculated using the HELAS
library\cite{HELAS}. The final state $\tau$ leptons are generated
using the BASES/SPRING package\cite{BASES}, and are decayed with 
TAUOLA version 2.3\cite{TAUORA}.  The effects of initial state
radiation, beam energy spread, and beamstrahlung are also taken into
account\cite{TOP}.

The end-product stable particles ($\pi^{\pm}$, $\gamma$, $e$, $\mu$... 
)  are then processed through a detector simulator, and are
identified, if possible, as ($\pi^{\pm}$ ,$\gamma$...) candidates. In
this paper, we assumed the JLC1 detector parameters, except
for the forward electron veto system. The model detector is
equipped with a central drift chamber(CDC), electromagnetic and hadron
calorimeters(EMC, HDC), and muon drift chambers, whose parameters can
be found in Ref.\cite{JLC1}. The used detector simulator is the same
as the one used in the previous studies\cite{TSUKA}.

The event signatures for the $\stau$ pair production are acoplanar two
jets or one jet + one lepton. The former mode is cleaner since the
latter mode suffers from $WW$, $e\nu W$, and $eeWW$ backgrounds. We
will, therfore, concentrate on the former mode. We used the following
basic cuts to select such an acoplanar two-jet event:

\begin{enumerate}
\item There exist two and only two jets for some $y_{\rm cut}>
2.5\times 10^{-3}$ where $y_{\rm cut}$ is imposed on  the reduced jet
invariant mass: $E_1E_2\cdot(1-\cos\theta_{12})/(E_{\rm vis})^2>y_{cut}$.
\item Both of the two jets must clear the polar angle cut $\vert
 \cos\theta_{\rm jet}\vert<0.8$.
\item The net charge of each jet must be unity and opposite in sign to
that of the other.
\item The acoplanarity of the two jets has to be large enough:
$\theta_{\rm acop}>30^{\circ}$.
\item These two jets have to have invariant masses consistent with
$\tau$ hypothesis: $m_{\rm jet1}$, $m_{\rm jet2} <3$ GeV.
\item The missing transverse momentum ($\sla{P}_T$) has to exceed 15 GeV.
\item There has to be no electron or position above $\theta_e>50$ mrad
from the beam axis.
\end{enumerate}

In addition we need cuts to identify $\tau$ decay products as $ \rho$
or $a_1$ in order to analyze each decay mode separately:

\begin{itemize}
\item $\rho$ cuts\\ A jet with two $\gamma$- and a 
$\pi^{\pm}$-candidates is identified as a $\rho^{\pm}$ 
$\rightarrow$ $\pi^{\pm}\pi^0$ candidate if $m_{2\gamma}<0.25$ GeV 
and $m_{\rm jet}<0.95$ GeV.\\ If there is only one $\gamma$ 
candidate in the jet ($\gamma+\pi^{\pm}$), we require 
$m_{\rm jet}<0.95$ GeV, assuming a possible cluster overlapping in 
the calorimeter.
\item $a_1$ cuts\\ A jet is identified as $a_1$ if it contains three 
charged $\pi$'s only, or four or three $\gamma$'s + one  
$\pi^{\pm}$, or two $\gamma$'s + one $\pi^{\pm}$ with  
$m_{2\gamma}>0.25$ GeV.
\end{itemize}

Cuts 1) through 4) are similar to the one used in the previous  
studies\cite{TSUKA}. These cuts together with cut 5) were designed to 
reduce  the background from 
gauge boson productions ($WW,e\nu W, eeWW$) to less than 1 fb 
for $P_{e}=0.95$ at $\sqrt{s}=500$ GeV.

Cuts 2),4), 6) and 7) are to remove the $e^+e^-\rightarrow
e^+e^-\tau^+\tau^-$ background, where the two photons radiated off
from the initial-state $e^+$ and $e^-$ collide to produce a $\tau$
pair, while the $e^+$ and $e^-$ escape into the beam pipe. Most of
proposed LC detectors have a relatively big acceptance hole in the
forward region. For example the JLC1 model detector can only veto
$e^{\pm}$'s above $\theta>150$ mrad, allowing the two-photon
background with a $\sla{P}_T$ up to as high as 75 GeV kinematically,
and up to 37.5 GeV typically, since it is quite rare that the two
initial-state particles give the maximum possible transverse kick in
the same direction to the $\tau^+\tau^-$ system.

In the previous Monte Carlo studies of the backgrounds for sfermion
productions, the $e^+e^-l^+l^-$ backgrounds were eliminated by cuts on
acoplanarity angle and $\sla{P}_T$\cite{DESY}.  The same applies to
the $ee\tau\tau$ background in the $\stau$ production studies. One might
worry about the $\tau$ decay giving extra $\sla{P}_T$ to 
the $ee\tau\tau$ events, however, the overall 
reduction of the energy of the $\tau$
decay products compensates this effect. The $\sla{P}_T>35$ GeV cut together
with an electron veto angle of $\theta_{e}^{veto}=150$ mrad, 
the cuts on polar angle (cut 2), and acoplanarity angle (cut 4) 
 turns out to remove most of the background events. 
We have generated 110K
$e^+e^-\tau^+\tau^-$ events with $\theta_e<150$ mrad,
$E_{\tau_1}+E_{\tau_2}>$15 GeV, $\vert\cos\theta_{\tau}\vert<0.9$, and
$\theta_{\rm acop}>10^\circ$, using a code developed by
Kuroda\cite{KURODA}.\footnote{We have generated the
$e^+e^-\tau^+\tau^-$ events in the phase space 
suffiently larger than the one defined by 
the cuts 2) and 4), because the reconstructed jet
axes  do not in general coincide with the original $\tau$ directions.}
The corresponding production cross section is 1.10 pb, and therefore
the generated events are about $\int{\cal L}dt=100$fb$^{-1}$ equivalent.
12 events survived cuts 1)-5), 7) and $\sla{P}_T>35$ GeV:
$\sigma(ee\tau\tau)\vert_{\rm cut}=0.12$fb$\pm 0.035$fb.

Introducing such a high missing $P_T$  cut might 
introduce an extra correlation
between the acceptance (or the measured value of the $\stau$ pair
production cross section) and the polarization of $\tau$ lepton from
$\sti$ decay, because jet energies become softer for a smaller $\ptau$
as discussed already in Sec.2.3.  Our MC simulation for
$\msti=150$ GeV, $\mchi=100$ GeV, and  $\sqrt{s}=500$ GeV, shows
that 17.3\%($P_{\tau}$=1) /17.1\%($P_{\tau}=-1$) of the generated signal
events are identified as $\sti$ with no $\sla{P}_T$ 
 but $E_{\rm vis}>10$ GeV
cut\footnote{The acceptance is smaller than that of the other sleptons
since we had to require both of the $\tau$'s decay hadronically. The
background to the search mode where one $\tau$ decays into $e/\mu$ is
larger but expected to be manageable. We also had to apply a tighter
jet polar angle cut to reduce the $e^+e^-\tau^+\tau^-$ background.},
while only 12.6\%($P_{\tau}$=+1)/9.8\%($P_{\tau}=-1$) of them
identified for $\sla{P}_T >$35 GeV.\footnote{The $P_{\tau}$ dependence comes
mostly from $\sti\sti\rightarrow\pi\rho$ events, where $\pi$'s tend to
have low energy for $P_{\tau}=-1$. Those events are less likely to be
accepted due to the $\sla{P}_T$ cut.} Though the correlation between
$P_{\tau}$ and the acceptance is smaller than that of the $\pi$ 
mode, and $P_{\tau}$ can also be constrained from the $z_c$ distribution of
the $\rho$ decays, the reduction of the acceptance by up to a factor of
2, and its strong $P_{\tau}$ dependence might be worrisome. (See Fig.5 for
the $\sla{P}_T$ distribution of the signal events corresponding to $10^4$
generated $\stau$ pairs and the $ee\tau\tau$ background for $\int{\cal
L}dt=100$fb$^{-1}$.) It should also be noted that events with smaller
jet energies are less likely to be accepted. This might complicate the
simultaneous measurement of $\mchi$ and $\mstau$, as the measurement
of the energy distribution near $E^{\tau}_{\rm min}$ becomes more
difficult (see Fig.6).

In this paper, we therefore assume a forward coverage down to
50 mrad. The $e^+e^-\tau^+\tau^-$ production cross section for
$E_{\tau_1+\tau_2} >15$ GeV, $\vert\cos\theta_\tau\vert<0.9$,
$\theta_{\rm acop}>10^{\circ}$, and $\theta_e<50$ mrad is 0.719 pb. Out
of 70K generated $e^+e^-\tau^+\tau^-$ events , which correspond to
$\int{\cal L}dt=100$fb$^{-1}$, only 19 events remained as background after
applying cuts 1-7). The overall detection efficiencies for the signal
events after the same cuts are 17.0\% and 16.0\% for $P_{\tau}=1$ and
$P_{\tau}=-1$, respectively. Jet energy distributions of
$\rho$-identified events for different $\sla{P}_T$ cuts are shown in Fig.6
for $\msti=150$ GeV, $\mchi=100$ GeV, and $P_{\tau}=-1$.\footnote{The jet
energy distributions are slightly softer than that of Fig.4d), as the
MC simulation includes beam effects and initial state
radiation. These effects will also be included in the fits of Sec.3.2.} 
One can see that the $\sla{P}_T>15$ GeV cut has no significant effect on the
signal events as expected, while for the $\sla{P}_T>35$ GeV cut, the acceptance
diminishes drastically for $E_{\rm jet}<50$ GeV, making the determination
of $E_{\rm min}^{\tau}$ difficult.

In order to realize the 50 mrad veto angle, we need to place
additional veto counters in the beam background mask, which might
produce extra beam backgrounds. On the other hand, having a tighter
forward veto can significantly reduce the SM background in the low
$\sla{P}_T$ region, which will help us extend our discovery reach to SUSY
particles with a mass which is very close to that of the LSP.  Further
studies are necessary to optimize the parameters for the extra forward
electron veto. Another possibility to reduce the $\sla{P}_T$ cut value is of
course to go down close to the $\stau$ pair production threshold. If
the $\stau$ production is accessible at $\sqrt{s}$=350 GeV, $\sla{P}_T>25$
GeV must be enough to eliminate the $ee\tau\tau$ background.

Now we are going to discuss the $\rho$ and $a_1$ cuts. These cuts are
chosen to minimize contaminations of $\rho$ to $a_1$ or vice versa due
to mis-reconstruction of photons. As described earlier, the $\rho$
and $ a_1$ decays involve $\pi^0$'s, which in turn decay into $2
\gamma$'s. For a high energy $\pi^0$, however, the two photons are
occasionally misidentified as a single photon due to the cluster
overlapping in the calorimeter, and therefore the $a_1$ sometimes has
the same signature as $\rho$. Fig.7a) shows the jet invariant mass
distributions of the events consisting of a $\pi^-$ and one or two
photon candidates coming from $\stau^+\stau^-\rightarrow
\tau^+\tau^-\rightarrow \pi^+\rho^-$ and $\pi^+a_1^-$. The solid
histogram is for $\rho^-$ decays from 50K $\stau$ pairs forced to
decay into $\pi^+\rho^-$, and the bars are the data of 7190 $\stau$
pairs forced to decay into $\pi^+a_1^-$. The $\pi^+\rho^-$ data are
scaled so that its relative normalization to $\pi^+a_1^-$ is
correct. Due to the mis-reconstruction the events have a considerably
smaller jet invariant mass distribution compared to that of $a_1$'s
decaying into $2\pi^-\pi^+$(dotted histogram). The $\rho$ cut on the
invariant mass $m_{\rm jet}<0.9$5 GeV only removes half of the $a_1$
contamination.

In Figs.7b) and c), we plot the jet energy and $z_c$ distributions of
the $\rho$ candidates that satisfy the $\rho$ cuts in the same
$\pi^+\rho^-$ and $\pi^+a_1^-$ samples. Solid histograms are those
from the $\pi^+\rho^-$ sample and bars are from the $\pi^+a_1^-$
sample. The number of identified $\rho$ events from
$\pi^+\rho^-$/$\pi^+a_1$ samples is 4522/400 for $P_{\tau}=+1$. The
contamination is larger for a higher $E_{\rm jet}$ and a lower $z_c$.
This is because high energy $\pi^0$'s from $a_1$ decays have less
chance to be identified as 2 photons, thereby sneaking into the $\rho$
signals. The same MC simulation told us that very few $a_1$ decays
could be reconstructed with $N_{\gamma}\ge 3$ if $E_{\rm jet}>50$ GeV.

The contamination affects $\msti$ and $P_{\tau}$ fit to $E_{\rm jet}$ and
$z_c$ distribution.  In principle the $a_1$ contamination to the
$\rho$ sample must be corrected for before the data are fitted to
obtain $m_{\stau}$ or $P_{\tau}$. However, due to rather low expected
statistics ($\sim$ 1400 $\rho$ candidates expected to survive after
the cuts for $10^4$ $\stau$ pairs), we did not attempt making such
corrections at all. We will see in the next subsection that input
parameters of MC and the corresponding best fit values of 
the $m_{\stau}$ and $P_{\tau}$ fits to ${\cal O} (10^4)$
$\sti$ pair events are consistent with each other.

\subsection*{3.2 Fit to MC data}

In this subsection we present the results of our fits to the selected
MC data for representative sets of parameters, with $\msti=150$ GeV,
$m_{\tilde{\chi}^0_1}=100$ GeV, and $\sqrt{s}=500$ GeV, and
backgrounds for $\int{\cal L}dt =100$ fb$^{-1}$ and $P_{e}=0.95$.

In sections 2.1 and 2.2, we have discussed the importance of measuring
$\msti$ and $\theta_{\stau}$ to determine the weak scale $\stau$ mass
matrix, and $P_{\tau}$ to determine $\tan\beta$. These parameters can
be measured by looking at $\rho$ candidates from $\sti$ cascade decay
(Sec.2.3): $\msti$ is measured through the energy distribution,
$\sin\theta_{\stau}$ through the production cross section
$\sigma_{\sti\sti}$, and $P_{\tau}$ through the $z_c$ distribution.

In the following, we fit the $E_{\rm jet}$ and $z_c$ distributions of
the $\rho$ candidates selected from the signal MC data to numerical
functions calculated by convoluting the $\tau\rightarrow\rho_{L(T)}$
decay spectra with the $\tau$ energy distributions.
The fit parameters
are $\msti$, $m_{\tilde{\chi}^0_1}$, $P_{\tau}$, and the number of
produced $\sti$ pairs $N_{\sti\sti}$. The results of the fit to $10^4$
and $5\times10^3$ $\sti\sti$ pairs will be shown in this
section. Notice that the production cross section of $\str$($\stl$)
with $m_{\stau}=150$ GeV is about 0.09 pb (0.02 pb) for $P_e=1$;
therefore, the generated $10^4$ $\sti$ events roughly correspond to
$\int{\cal L}dt =100 \ (400)$fb$^{-1}$, respectively (see Fig. 1 of
Ref.\cite{NOJIRI}). The fit will be extended to include  
the measurement of $\se_R$ production and decay in Sec. 4, to 
obtain the error on $\tan\beta$.

We first describe our calculation of the theoretical distributions for
the fit.  As we mentioned earlier, the $E_{\rm jet}$ and/or $z_c$
distributions are numerically calculated by convoluting the
$\tau\rightarrow\rho_{L(T)}$ and the $\tau$ energy 
distributions.\footnote{
We included the effect of the finite $\rho$ width for 
the jet energy distribution 
as in Ref.\protect\cite{HAGI} but did not take it into account for the 
$z_c$ distribution.}  In
the calculations of the theoretical distributions, we took into
account the effect of beamstrahlung and initial state radiation  on the 
$\sti$ energy distribution as before\cite{TOP}.  We also found that
the acoplanarity angle cut (cut 4) has a significant effect on the
energy distribution.  Since the acoplanarity angle cut is very
complicated to implement in the numerical calculation of the energy
distribution, we approximated the effect by imposing an acolinearity
cut of 30 degrees in the CM frame of the $\sti$ pairs instead.  The
resultant jet energy distribution that was calculated this way 
roughly reproduced the shape of the energy distribution of a
statistically larger MC event sample ($10^5$ $\sti$ pairs).  The
overall normalization has been determined by comparing it with the MC
simulation, and corrected for a $P_\tau$-dependent acceptance
factor. Agreement between the $z_c$ distribution of the MC data and
that of the numerical calculation was poor for $z_c\sim 0$ or $z_c\sim
1$ due to the acceptance effects; we therefore fit the $z_c$
distribution only over the range $0.08\leq z_c\leq 0.92$.

In the previous subsection we have seen that the $\sla{P}_T>15$ GeV cut is
necessary to reduce the $ee\tau\tau$ background for an electron veto
of 50 mrad.  Since the signal events are hardly affected by the cut,
and since it is hard to implement the $\sla{P}_T$ cut in the numerical
calculation of the fitting curve, we decided to ignore the $\sla{P}_T$ cut
for both the MC events and the fitting curve, and neglected the
$ee\tau\tau$ background altogether, though the dominant backgrounds
from $WW$, $eeWW$, $ZZ$, and $\nu\bar\nu Z$ corresponding to
$\int{\cal L}dt =100$ fb$^{-1}$ have been included in the fits in this
section.\footnote{$E_{\rm vis}>10$ GeV is implicit  for all the MC
event generation  in this subsection. This condition is not included in the
fitting curves as its effects were found negligible.} The production
cross sections for the backgrounds before and after the 
selection cuts and the
number of remaining $\rho$ events are listed in Table 1.

We first separately perform a fit of $\msti$ and $\mchi$ to the $E_{\rm
jet}$ distribution, and that of $P_{\tau}$ to the $z_c$ distribution.
Figs.8 a) and b) are the results of our mass fit to the $E_{\rm jet}$
distribution of $10^4$ $\sti\sti$ events decaying into $\tau_R
\tilde{\chi}_1^0$ exclusively. 1476 events were identified as $\rho$
and used for the jet energy fit.  In this fit, we kept $P_{\tau}=+1$
and set the normalization of the curve so that the total number of
events agreed with that of the MC data. Fig.8a) plots the jet energy
distribution of the MC events together with the best fit curve
obtained by minimizing the log-likelihood function $(\equiv \chi^2)$
with $\msti$ and $\mchi$ as free parameters. The contours for
$\Delta\chi^2=1$ and $4$ in $\mchi$-$\msti$ plane are shown in Fig
8b). The MC events were generated for $\msti=150$ GeV and $\mchi=100$
GeV, while the best fit values are $\msti=147$ GeV and $\mchi=97$ GeV
for this MC sample. The values are consistent with the inputs;
$\Delta\mchi$=2.8 GeV and $\Delta\msti=3.9$ GeV can thus be expected
as 1-$\sigma$ errors on these quantities.

The errors on the two parameters might be reduced further.  Notice
that we have only used the events identified as
$\sti\rightarrow\tau\rightarrow\rho$ here. The other modes into $\pi$,
$a_1$, or other leptons can also be used to increase the
statistics. One may combine the information from other sparticle
decays once they are observed. The previous analysis of $\smu$ decays
showed that $\Delta\mchi=1$ GeV can be achieved typically, which would
reduce $\Delta\msti$ down to 1.5 GeV.

On the other hand, $P_{\tau}$ can be determined by fitting the
$\pi^{\pm}$ fraction of the parent $\rho$ energy ($z_c\equiv
E_{\pi^\pm}/E_{\rm jet}$). In this fit, we fixed $\mchi$ and $\msti$
to their input values and the normalization of the fitting curve to
the total number of the MC sample. Figs.9a) and b) show the $z_c$
distribution for the selected $\rho$ candidates together with the best
fit curve. Here we used the data in the region $0.08\leq z_c \leq
0.92$ and $E_{\rm jet}>20$ GeV since the low energy region is
insensitive to $P_{\tau}$. 924(885) MC events were used for the fit
where $\sti$'s decayed into $\tau_R(\tau_L)$'s.  The best fit values
and their errors were obtained to be $0.995\pm0.082$ and $-0.991\pm
0.08$ for $\tau_R$ and $\tau_L$, respectively.

In order to justify such separate fits of $E_{\rm jet}$ and  
$z_c$ distributions
with some of the fitting parameters fixed by hand, we must make
sure that there is no strong correlation among $\msti$, $\mchi$,
$P_{\tau}$, and $\sigma_{\sti\sti}$. For this purpose, we calculated
the errors on the masses and $P_{\tau}$ by fitting a two-dimensional
distribution in $(E_{\rm jet}, z_c),$ varying $\mchi$, $\msti$,
$P_{\tau}$, and the total number of produced stau pair events
$N_{\tau\tau}$ (which corresponds to $\sigma_{\sti\sti}$ if the
integrated luminosity and acceptance  is known). 
1224 events in the interval $0.08\leq
z_c\leq 0.92$ were used for the fit with $P_{\tau}=1$.

The resultant errors obtained from this fit agreed very well with the
previous estimates.  The best fit value for the masses are
$\msti=146.3$ GeV and $\mchi=95.4$ GeV. The shape of the $\chi^2$
contour projected onto the $\mchi$-$\msti$ plane looked quite similar
to that of Fig.8. The estimated errors are $\Delta\msti=4.07$ GeV and
$\Delta\mchi=2.99$ GeV, being consistent with the previous estimates,
taking into account the difference in the number of events used for
each fit. The error on $P_{\tau}$ was also calculated allowing
$\mchi$, $\msti$, and $N_{\tau\tau}$ to move freely in minimizing
$\chi^2$. The best fit value of $P_{\tau}$ is $P_{\tau}=0.89\pm 0.07$,
again consistent with the previous estimate.  These results support
our assumption of small correlation between the energy distribution
and $P_{\tau}$ for the $\tau\rightarrow\rho$ decay mode.

Finally we move on to the determination of the $\stau$ mixing angle
$\theta_{\stau}$. As discussed already, our strategy is to use the
measurement of the production cross section together with that of
$m_{\tilde\tau}$. We generated 5000 $\sti$ pairs decaying into
$\tau\tilde{\chi}^0_1$ with $P_{\tau}=0.6788$. The SM backgrounds for
$\int{\cal L}dt =100$ fb$^{-1}$ were also included. The $\rho$ signal
sample corresponds to $\sin\theta_{\stau}=0.7526$, and $\msti=150$ GeV
with a bino dominant LSP.

We used events in the region $0.08<z_c<0.92$, which, after selection,
reduced to 628 events. Since the acceptance differs by about 12\%
between $P_{\tau}=1$ and $P_{\tau}=-1$ due to the $z_c$
cut,\footnote{The data in $z_c\sim 0$ or $1$ can of course  be used
once detector performance is understood and included in the numerical
calculation of fitting curves. The dependence of the acceptance to
$P_{\tau}$ described here is purely artificial, unlike that caused by
$\sla{P}_T$ cuts.} we minimized $\chi^2$ in the $N_{\sti\sti}$- $\msti$
plane, varying $P_{\tau}$, $\msti$, and $\mchi$. No significant
correlation was found between $N_{\sti\sti}$ and $\mchi$ and the
estimated errors are $\Delta\msti=6.6$ GeV and
$\Delta\sigma_{\sti\sti} =2.2$fb.\footnote{We assumed $\int{\cal
L}dt=100$ fb$^{-1}$.} The mass error is consistent with the
previous estimate if one takes into account the difference in the
numbers of produced events and the acceptances.  The error on the cross
section is consistent with the error simply estimated by the
statistics of the accepted events.

Fig.10 plots contours of constant minimized $\chi^2$ surfaces
projected onto the $\msti$-$\sin\theta_{\stau}$ plane. We found
$\Delta\sin\theta_{\stau}=0.049$. One can see that the correlation
with $\msti$ makes the error large. It is possible to reduce the
$\sti$ mass error by using the $\mchi$ obtained from other
measurements as it was discussed earlier. $\Delta\mchi=1$ GeV would
reduce $\Delta\msti$ to 2.21 GeV, in which case the error on
$\sin\theta_{\stau}$ is less than 0.03. However, the error cannot be
less than 0.014, which is limited by the observed number of 
events.\footnote{Including other decay mode to the analysis
would improve statistice, and reduce $\Delta\theta_{\tau}$.}

The $\sti$ decay process studied above is quite complicated compared
to that of $\se$ or $\smu$. Nevertheless, in the above discussions, we
have found the measurements of masses $\msti$ $\mchi$, $P_{\tau}$, and
$N_{\sti\sti}$ could be done without any significant correlations each
other. We have also learned the mass errors using ${\cal O}(600)$
accepted $\rho$ events are consistent with those of ${\cal O}(1400)$
events if the latter is statistically scaled. The error on the cross
section is consistent with the error estimated by the statistics of
the accepted events. This allows us to estimate the errors on $\msti$,
$P_{\tau}$, $N_{\stau\stau}$, and $\theta_{\stau}$ reliably by simple
statistical scaling of each error in a wide region of parameters
space. This fact is used when we combine the $\sti$ measurements with
the $\se$ measurement in the Sec. 4.

In this subsection we assumed that a  tight forward electron veto
is possible and applied a  small $\sla{P}_T$ cut as was discussed in the
previous subsection. If this $\sla{P}_T$ cut has to be increased to 35 GeV,
the result of this subsection must change. The mass measurement is
based on the measurement of $E^{\tau}_{\rm max(min)}$ extracted from
the energy distribution of $\rho$ candidates. As can be seen from
Fig.7, events near $E^{\tau}_{\rm max}$ are not affected by the cut,
therefore $\Delta E^{\tau}_{\rm max}$ will not change,
either. $E^{\tau}_{\rm max}$ is sensitive to $\msti-\mchi$, therefore
$\Delta\msti$ obtained with $\mchi$ from other slepton measurements
will not change significantly. $\Delta P_{\tau}$ will not be affected
too much, either, since the $z_c$ distribution is not sensitive to
$P_{\tau}$ for $E_{\rho}\lesssim 20$ GeV, which is the region most
affected by the $\sla{P}_T>35$ GeV cut. Finally, some extra dependence of
the acceptance on $P_{\tau}$ should be introduced by the large $\sla{P}_T$
cut. This might increase the error on the production cross section,
since the acceptance moves by 20\% with $P_{\tau}$ varying from $1$ to
$-1$. However, the dependence can be tamed by measuring $P_{\tau}$
from the $z_c$ distribution.

\section*{4. Combined Analysis}
\subsection*{4.1 $\dbarchi_{\se}$ and  $\dbarchi_{\stau}$  functions}
In this section we are going to extract $\tan\beta$ by combining the
measurements of $\ptau$, $\sin\theta_{\stau}$, and the knowledge of
the neutralino mass matrix obtained from the measurements of $\se_R$
production and decay. Some MC simulation of $\tilde{e}$ production had
already been carried out for a specific set of parameters\cite{TSUKA},
and we have just finished a corresponding MC analysis for $\sti$ in
the previous section. It is now straightforward to perform combined
fits to determine the MSSM parameters for representative points in the
parameter space.  Nevertheless it is quite time consuming to do it
exactly as in a real experiment.

Therefore in this paper, we define $\Delta\chi^2$-like function
$\Delta\bar{\chi}^2$ to estimate the sensitivity of LC experiments to
these SUSY parameters. As we mentioned already, we take a sample case
where both $\sti$ and $\tilde{e}_R$ are produced at a future LC. The
$\tilde{e}_R$ pairs are selected requiring acoplanar $e^+e^-$ paris,
while the $\sti$ pairs are taken from an acoplanar two-jet
sample. Since these samples are statistically independent of each
other, we define the $\tilde{e}$ and $\sti$ parts of the
$\Delta\bar{\chi}^2$ functions $\dbarchi_{\se}$ and $\dbarchi_{\stau}$
separately in this subsection.  $\dbarchi_{\se}$ and
$\dbarchi_{\stau}$ are functions of two sets of input parameters:
$(m_{\tilde{e}_R}, M_1(M_2), \mu, \tan\beta) $ and
$(m'_{\tilde{e}_R},M_1'(M_2)', \mu', \tan\beta')$ for
$\dbarchi_{\se}$, and $(\msti, \theta_{\stau},M_1(M_2), \mu,
\tan\beta )$ and $(\msti', \theta'_{\stau}, M'_1(M'_2), \mu ',
\tan\beta')$ for $\dbarchi_{\stau}$. The $\dbarchi$ function are
defined in such a way that $\dbarchi=0$ when the two sets are equal
and the projection of the hypersurface of $\dbarchi=1(,4,9..)$ to one
of the parameters $(m'_{\se_R}, M_1'....)$ fixing $(m_{\se_R},
M_1...)$ roughly agrees with the 1(,2,3..)$-\sigma$ error of that
parameter.  In this subsection, we first define
$\dbarchi_{\se(\stau)}$ in detail, and then discuss the error on
$\tan\beta$ in the next subsection. Readers who are interested only in
the results can skip this subsection.

The polar angle($\theta$) distribution and the end point energies of
electrons from $\se_R$ decays can be measured at future LC
experiments as discussed already in Sec. 2.2. Therefore we define
$\dbarchi_{\se}$ by using the two sets of quantities as follows:
\beq\label{e12} \dbarchi_{\se}( m_{\tilde{e}_R}, &&M_1(M_2), \mu,
\tan\beta; m_{\tilde{e}_R}', M_1'(M_2)', \mu', \tan\beta')\nonumber\\
&&=\sum^{n_{bin}}_{i=1}\frac{(n_i'-n_i)^2}{n_i}
+\left(\frac{E^{e'}_{\rm max}-E^e_{\rm max}}{\Delta E^e_{\rm
max}}\right)^2 +\left(\frac{E^{e'}_{\rm min}-E^e_{\rm min}}{\Delta
E^e_{\rm min}}\right)^2, \eeq where $n_i$ and $n_i'$ are the expected
numbers of events in $i$-th bin between $-1 + 2(i-1)/n_{bin}\le$
$\cos\theta$ $< -1+2i/n_{bin}$ calculated for the first and the second
sets of input parameters: ($m_{\tilde{e}_R}$, $M_1(M_2)$, $\mu$,
$\tan\beta$) and ($m_{\tilde{e}_R}'$, $M_1'(M_2)'$, $\mu'$,
$\tan\beta'$), respectively.  For later use we calculated $n_i$
assuming $\int{\cal L}dt=20$ (or 100) fb$^{-1}$, $\sqrt{s}=500$ GeV, a
27 \% acceptance, and $n_{bin}=25$, making use of formulas for the
$\se_R$ pair production cross section listed in Appendix B. The
acceptance is chosen to be a factor of 0.6 smaller compared to the
value obtained by the MC simulation\cite{TSUKA}. This is because our
expression for the selectron production cross section does not include
any effects of initial state radiation, beam energy spread, and
beamstrahlung.

$E_{\rm max}^{e(')}$ and $E_{\rm min}^{e(')}$ are the upper and lower
end points of the energy distribution of electrons for the first(the
second) set of parameters.  $\Delta E^e_{\rm max}$ and $\Delta
E^e_{\rm min}$ are defined by \be\label{e13} \Delta E^e_{\rm
max(min)}\equiv \sqrt{\frac{2}{N_{\rm av}}}\times E_{bin}, \ee where $N_{\rm
av}=n_{\rm total}E_{bin}/(E^e_{\rm max}-E^e_{\rm min})$ with
$E_{bin}=4$ GeV being a kind of bin width, and $n_{\rm
total}=\sum_{i=1}^{n_{bin}}n_i$.

$\dbarchi_{\se}$ is chosen to reproduce the actual $\Delta\chi^2$ of
the MC data fitted with the second set of parameters
($m_{\tilde{e}_R}'$, $M_1'(M_2)'$, $\mu'$, $\tan\beta'$) when $n_i$,
$N_{\rm av}\gg 1$ and $n_{\rm total}\gg N_{\rm av}$.  The reason is
the following: If $n_i$'s are replaced by actual data, the first term
of Eq.(\ref{e12}) is $\chi^2$ of the data fitted with the parameters
($m_{\tilde{e}_R}'$, $M_1'(M_2)'$, $\mu'$, $\tan\beta'$) based on
Gaussian distributions.  The difference between the data and $n_i$'s
divided by $n_i$ must be small in the limit of large statistics if
($m_{\tilde{e}_R}$, $M_1(M_2)$, $\mu$, $\tan\beta$) is the parameter
set that nature has taken, therefore the projection of the
hyper-surface that satisfies $\dbarchi_{\se}=1$ to one of the fitting
parameters roughly indicates the size of $\pm 1$-$\sigma$ deviation of
the parameter from the best fit point.  In this sense, we can call the
first set of parameters as input parameters and the second set as
fitting parameters.

The second and the third terms are intended to represent the
sensitivity of the electron energy distribution measurement to
determine $E^e_{\rm max}$ and $E^e_{\rm min}$.  As has been mentioned
already, $m_{\tilde{e}_R}$ and $\mchi$ can be determined from the
endpoints of the energy distribution of electrons. In actual
experiments, electron energies are measured by some detector with a
finite energy resolution. The JLC1 model detector, for example, has
$\sigma_E/E= 15\%/\sqrt{E}\oplus 1\%$\cite{JLC1}, demanding us
to take into account the finite size of $E_{\rm bin}\sim 2$ GeV.
Moreover the energy distribution can also be smeared and distorted due
to finite beam energy spread, beamstrahlung, initial state radiation,
and possible dependence of acceptance on electron energies. In the
previous study the shape of the energy distribution and its dependence
on $m_{\tilde e}$ and $m_{\tilde{\chi}^0_1}$ were obtained from the MC
study itself, and the mass errors were estimated by actually fitting
the energy distribution.

For simplicity we assume here that the energy distribution is flat
between $E^e_{\rm max}$ and $E^e_{\rm min}$, while conservatively
taking $E_{\rm bin}$= 4 GeV. If the average number ($N_{\rm av}$) of
events in a single bin is large enough so that the fluctuation is
negligible compared to $N_{\rm av}$, the central value of $E_{\rm
max(min)}$ is obtained as
\be\label{e14} 
E^e_{\rm max(min)} = E_{\rm end}^c \pm
\left(\frac{N_{\rm bin}-N_{\rm av}/2}{N_{\rm av}}\right) E_{\rm bin}, 
\ee 
where $E_{\rm end}^c$ is the central energy of the upper (lower) edge
bin, $N_{\rm bin}$ is the number of events in it, and $N_{\rm av}$ is
the average number of events in some intermediate bin. Based on the
statistical error on $N_{\rm bin}$ estimated assuming a Gaussian
distribution, the error on $E^e_{\rm max(min)}$ is $\Delta E^e_{\rm
min(max)}/E_{\rm bin}=$ $1/\sqrt{N_{\rm bin}}$.

However, when the actual $E^e_{\rm max(min)}$ is very close to a bin
boundary, the fluctuation of $N_{\rm bin}$ becomes
non-Gaussian. $N_{\rm bin}$ can even exceed $N_{\rm av}$ or becomes
zero, making nonsense of Eq.(\ref{e14}) and the error estimate.
Therefore in Eq.(\ref{e13}), we replaced $N_{\rm bin}$ by $N_{\rm
av}/2$ , which corresponds to the choice of binning where $E^e_{\rm
max(min)}$ is approximately at the center of the edge energy bin.  In
such a case, the fluctuation of the edge bin becomes Gaussian-like as
long as $N_{\rm av}$ is large enough, thereby justifying our
estimation.

Several comments are in order. One might think that the measurement of
the end point energies does not fit any $\chi^2$ analysis implying a
Gaussian distribution, if the energy resolution is too good and the
expected numbers of the events in the edge bins are too small. In such
a case the probability distribution for $E^e_{\rm max}$ or $E^e_{\rm
min}$ is expected to be asymmetric, because if an event is observed in
some energy bin between $E_1$ and $E_2$, selectron and neutralino
masses which give $E^e_{\rm max}<E_{1}$ or $E_2 < E^e_{\rm min}$ are
strongly disfavoured. However, the number of events in a single bin is
expected to be large enough in the ``precision measurement'' phase of
the LC , and hence our treatment assuming a Gaussian distribution can
be justified: we have checked if our treatment of $\Delta E^e_{\rm
max(min)}$ roughly reproduces the previous results on $\tilde{\mu}$
production and decay\cite{TSUKA}, and found that the $\dbarchi$
contours by using the last two terms of Eq.(\ref{e12}) but replacing
$E^e_{\rm max(min)}$ to $E^{\mu}_{\rm max(min)}$ in the
$m_{\tilde{\mu}}$-$m_{\tilde{\chi}^0_1}$ plane agree very well with
the previous results for the same number of accepted events.

When we calculated the $\theta$ distribution or the electron energy
distribution we occasionally found $n_i$ less than 15, or $3\Delta
E^e_{\rm max(min)}>E_{\rm bin}/2$. In such a case, we merged the
$\cos\theta$ bins or enlarged $E_{\rm bin}$. Our treatment
underestimates the sensitivity compared to any log-likelihood
analysis based on a Poisson distribution and therefore is
conservative.

The $\dbarchi$ analysis mimics the true $\chi^2$ fit to the $\theta$
and $E_{e}$ distributions, though it neglects the correlation between
the $\theta $ and $E_e$ distributions through the total number of
events: the fluctuations of the events in $E_e$ bins have correlations
with the fluctuations of $n_i$'s, since the events must add up to an
equal number in both distributions.  This correlation disappears,
however, in the limit where the number of events in the edge $E_e$
bins is negligible compared to the total number of events, thereby
justifying our method.

Finally, in the definition of $\dbarchi_{\se}$, we assumed that the
$\tilde{e}$ production angles are reconstructed precisely. This is not
true since there is always a wrong solution of $\theta$ for each
event. The wrong solutions must first be included in the $\theta$
distribution, which must then be subtracted statistically, bringing
more uncertainty into our analysis. We also assumed that all the
selected selectrons contribute to the determination of the production
cross section, angular distribution, and masses, and will not distort
the measurement due to $\tilde{e}_R$ decays into heavier neutralinos.

For $\sti$ pair production and their cascade decay $\sti\rightarrow
\tau\rightarrow\rho$, we have already discussed that it is important
to measure the total cross section, $P_{\tau}$ from the decay
distribution of the $\rho$ decay products, and $E^{\tau}_{\rm min}$
and $E^{\tau}_{\rm max}$ from the energy distribution of $\rho$
candidates. Fits to $10^4$ $\sti$ pairs have been done in Sec. 3.2,
taking into account backgrounds corresponding to $\int{\cal L}dt=100$
fb$^{-1}$, and the errors on $\msti$, $\mchi$, and $N_{\sti\sti}$ have
been obtained. In the following we define the $\sti$ part of our
$\Delta\chi^2$-like function $\dbarchi_{\stau}$ so that it reproduces
the results in Sec.3.2:
\be\label{e15}
\dbarchi_{\stau}=\frac{(N-N')^2}{N}+ \left(\frac{E^{\tau}_{\rm
max}-E'^{\tau}_{\rm max}}{\Delta E^{\tau}_{\rm max}}\right)^2
+\left(\frac{E^{\tau}_{\rm min}-E'^{\tau}_{\rm min}}{\Delta
E^{\tau}_{\rm min}}\right)^2 +\left( \frac{P_{\tau}-P'_{\tau}}{\Delta
P_{\tau}}\right)^2, 
\ee 
where $N^{(')}$ is defined to be the sum of  constant background
($N_{\rm bg}$) and the total number of signal $\rho$ events
($N^{(')}_{\rm total}$) for which  both of $\tilde{\tau}$'s decay directly
into $\tilde{\chi}^0_1\tau$ and the $\tau$'s then decay hadronically.
We took $N_{\rm bg}=100$. $N_{\rm total}^{(')}$ was estimated
using an integrated luminosity of 100 fb$^{-1}$, the tree level cross
section without any beam effects, and the acceptance obtained in the
previous simulation with no $\sla{P}_T$ cut. The branching ratio to $\tau
\tilde{\chi}^0_1$ was calculated by the formula in Ref.\cite{NOJIRI}. 
Notice that in the region where the lightest neutralino is
higgsino-like, the lighter chargino $\tilde{\chi}^-_1$ and the
lightest and the second lightest neutralinos $\tilde{\chi}^0_1$ and
$\tilde{\chi}^0_2$ are almost mass-degenerate. Then the decay modes
$\sti\rightarrow \tilde{\chi}^0_2\tau$ and
$\sti\rightarrow\tilde{\chi}^-\nu_{\tau}$ generally open up, which
would yield rather complicated final states with associated jets. As
we have not studied the sensitivities and backgrounds to these modes,
we will not include them in the study below, instead we will simply
take the number of events where both $\sti$'s decay into
$\tilde{\chi}^0_1$ to estimate $N_{\rm total}^{(')}$.

The first term of Eq.(\ref{e15}) is intended to show the statistical
significance of the total $\sti$ pair production cross section. On the
other hand the second and the third terms express sensitivity to
$E^{\tau}_{\rm max}$ and $E^{\tau}_{\rm min}$.  We again calculate
$\Delta E^{\tau}_{\rm max(min)}$ using a rather simple set of formulas:
\be\label{e16} \Delta E^{\tau}_{\rm max}=4.8 \sqrt{2/N_{\rm av}},\ \ \
\Delta E^{\tau}_{\rm min}=1.8 \sqrt{2/N_{\rm av}}, \ee where $N_{\rm
av}=E_{\rm bin} N_{\rm total}/(E^{\tau}_{\rm max}-E^{\tau}_{\rm min})
$. The effect of the smearing of the energy distribution by the
cascade decay of $\sti$ is taken into account by the overall factors
in the right hand side of Eq.(\ref{e16}).  The factors are chosen so
as to reproduce $\Delta{\msti}$ and $\Delta{\mchi}$ in the previous
subsection. The larger factor for $\Delta E^{\tau}_{\rm max}$ compared
to $\Delta E^{\tau}_{\rm min}$ may be understood as the effect of the
higher reduction of the events near $E^{\tau}_{\rm max}$ due to the
$\tau$ decays( see Figs. 4c), and d)). Finally, $\Delta P_{\tau}$ is
estimated statistically scaling the error on $P_{\tau}$ in Sec.3.2:
\be\label{e17} 
\Delta P_{\tau} = 0.07\times \left( 1400/\sqrt{1400+N_{\rm bg}}\right) 
/\left(N_{\rm total}/\sqrt{N}\right).
\ee

In Eq.(\ref{e15}), we assumed that there is no large correlation among
the measurements of $N_{\sti\sti}$, $E^{\tau}_{\rm max(min)}$, and
$P_{\tau}$. We also imply that errors on these parameters can be
estimated by the statistical scaling of the results in Sec.3.2. These
features have been checked explicitly by the MC analysis in the same
subsection.

\subsection*{4.2 Determination of $\tan\beta$ from slepton production}

We have already pointed out in Sec.2.2 that the simultaneous
measurements of $\sti$ and $\tilde{e}_R$ productions would determine
$\tan\beta$. In order to estimate its statistical error expected at a
future LC, we have defined $\dbarchi_{\se(\stau)}$ functions in the
previous subsection. $\dbarchi_{\se}$ is a function of
two sets of MSSM parameters $(m_{\se_R},M_1....)$ and
$(m'_{\se_R},M_1'....)$.  In the limit of infinite statistics, the
projection of the hypersurface of $\dbarchi=1(,4,9..)$ to one of the
parameters $(m'_{\se_R},M_1'....)$ fixing $(m_{\se_R}, M_1...)$ agrees
with the $1(,2,3..)$-$\sigma$ error of that parameter obtained by
using the $\cos\theta$ distribution and the end point energies of
electrons from $\se_R$ decays of real data. The definition of
$\dbarchi_{\tilde{\tau}}$ is similar, but here the data used are the
number of signal $\rho$ events from  $\sti\rightarrow$
$\tilde{\chi}^0_1\tau$ followed by  $\tau\rightarrow\nu_{\tau}\rho$, 
the end point energies of the $\tau$'s the $\sti$ decays, 
and the average $\tau$  polarization.

We will start our discussion with the determination of the parameters
of the neutralino mass matrix from $\tilde{e}_R$ production alone.
$m_{\tilde{e}_R}$ and $\mchi$ are determined essentially by 
through the energy distribution of the electrons from $\se_R$ decays.
On the other hand, $M_1$ is mainly constrained by the $\tilde{e}_{R}$
production cross section. The dependence of the total cross section on
$M_1$ and $\tan\beta$ for fixed $m_{\se_R}$ and $\mchi$, assuming the
GUT relation between $M_1$ and $M_2$, has been shown in
Fig.3a). Notice that constraining $m_{\tilde{e}_R}$ is very important
for the determination of $M_1$, as the production cross section
depends not only on $M_1$ but also on $m_{\tilde{e}_R}$.

Fig.11 shows the error on $M_1$ estimated with the $\dbarchi_{\se}$
function, where the input parameters were chosen such that
$m_{\tilde{\chi}^0_1}=100$ GeV, $P_{e}=+1$, $m_{\tilde{e}_R}=200$ GeV,
$\mu>0$, and $\int{\cal L}dt =20$ fb$^{-1}$. Both positive
and negative errors on $M_1$ are shown in \%.  The errors were
calculated by finding a minimum value of $\dbarchi$ for a fixed
$M_1'=M_1+\Delta M_1$ varying the other fitting parameters $(m'_{\se},
\mu', \tan\beta')$, with the MINUIT program. The values of $\Delta M_1$
which give $\Delta\bar{\chi}^2_{\rm min}=1$ are plotted as 1-$\sigma$
lines in the figure. This corresponds to projecting the
$\Delta\bar{\chi}^2_{\se}=1$ hypersurface to the $M_1$ axis.  $M_1$
will be determined within an error of $5\sim 7$\%, typically in the
region of parameter space shown in the figure.\footnote{We found that our
error on $M_1$ is smaller than previously quoted\protect\cite{TSUKA}. The
previous estimate did not use the GUT relation between $M_1$ and $M_2$
and the constraint on $\mchi$ which would have been obtained from the
$\tilde{e}_R$ production was not exploited in the $M_1$ and $M_2$
fit\protect\cite{PRIVATE}. Our result is therefore consistent with the
previous one.}

We found that the errors are asymmetric for positive and negative
fluctuation of $M_1$. This is because the condition $\dbarchi=1$
forces the total cross section to be close to its input value. The
cross section increases with decreasing $\tan\beta$ for $\tan\beta<5$
and decreases with increasing $M_1$ for fixed $\mchi$, as can be seen
in Fig.3. For $\tan\beta=1.5$, the error on $M_1$ is therefore larger
in the negative fluctuation (Fig.11): Increasing  $\tan\beta$ can
compensate the increase of $\sigma_{\stau\stau}$ due to the reduction of
$M_1$ here. For $\tan\beta=15$, a similar argument shows that the
error is larger in the positive fluctuation.

In the figure, we have also shown the error on $M_1$ for
$\tan\beta=15$, but restricting $\tan\beta'>10$ for searching minimum
value of $\dbarchi_{\se}$. In this case, the 1$\sigma$ fluctuation is
symmetric and smaller. This suggests that even a rough estimate of
$\tan\beta$ can greatly help us  restrict $M_1$. As we will see
below, such improvement is indeed possible in some region of parameter
space if $\sti$ production is observed.

Now we turn to the determination of $\tan\beta$. As we have discussed
already in Sec.2.2, $\tan\beta$ can be extracted from the polarization
of $\tau$ leptons produced in $\sti$ decays, if we know
$\theta_{\stau}$ and the neutralino mixing angles $N_{ij}$.

$\theta_{\stau}$ is determined from the measurement of $\msti$ and
$\sti$ production cross section. (The result from our  full MC analysis is
 in Fig.10 in Sec. 3.2).  The sensitivities to the production
cross section, $\msti$, and $P_{\tau}$ are taken into account in the
definition of the $\dbarchi_{\stau}$ as the first term, the second and
the third terms, and the fourth term of the right-hand side of
Eq.(\ref{e15}), respectively.

On the other hand some information on neutralino mixing can be
obtained from selectron production, e.g. by minimizing
$\dbarchi_{\se}$. The Neutralino mixing depends on only 3 (4) parameters
$M_1(M_2), \mu,\tan\beta$, where  $\dbarchi_{\se}$ 
strongly constrains two of
these three parameters, $M_1$ and $\mchi$, as  described earlier in
this subsection.

Notice that the constraint on $\mchi$ from $\dbarchi_{\tilde e}$ is
stronger than that from $\dbarchi_{\stau}$, because of the larger
statistics fot the  $\se$ production and the smearing effect of the end
point energies for $E^{\tau}_{\rm max(min)}$, which 
have been  taken into
account in the definition of $\Delta E^{\tau}_{\rm max}$ and $\Delta
E^{\tau}_{\rm min}$ in Eq.(\ref{e16}). The small $\Delta\mchi$ from
the $\se_R$ production helps  determine $\tan\beta$ better, for the
following reason: $\Delta\msti$ correlates with $\Delta\mchi$, and
$\Delta\theta_{\stau}$ with $\Delta\msti$, as can be seen in Fig.8b)
and Fig.10. Therefore $\Delta\theta_{\stau}$ is smaller for smaller
$\Delta\mchi$, which reduces the uncertainty coming from $\stau$
mixing in  determining $\tan\beta$ from $P_{\tau}$.

Fig.12 plots $\dbarchi=$ $\dbarchi_{\se}+\dbarchi_{\stau}=1$ contours
projected onto the $M_1$-$\tan\beta$ plane for input values of
($M_1$(GeV), $\tan\beta$)= (219.0,15),(149.3,15),
(124.5,15),(124.5,25), and (124.5,5). The other input parameters are
common for all representative points,$m_{\tilde{e}}=200$ GeV,
 $\msti=150$ GeV, $\mchi=100$
GeV, and $\sin\theta_{\stau}=1$ ($\sti=\tilde{\tau}_R$) and we took
$\int{\cal L}dt=100$ fb$^{-1}$.

As we have discussed previously, $P_{\tau}$ depends sensitively on
$\tan\beta$ if $\mchi\ll M_1$ or $\tilde{\chi}^0_1$ is higgsino-like,
because in this case the $\chi\stau\tau$ coupling involves the $\tau$
Yukawa coupling. Thus the error bar is expected to be smaller
for a larger $M_1$. However, the number of accepted events becomes
small when $\mchi\ll M_1$ and $\sti$ also decays into
$\tilde{\chi}^0_2$ or $\tilde{\chi}^{\pm}_1$, therefore the error on
$\tan\beta$ for $M_1=219$ GeV is larger than that of $M_1=149.3$
GeV. When $\tan\beta=15$, the lower(and upper) bounds of $\tan\beta$
for the $\dbarchi=1$ contours are 13.85 (18.5), 13.28 (16.37),
8.94 (18.74) for $M_1=219, 149.3$, and $124.5$ GeV, respectively.

Finally, in our definition of $\dbarchi$ we did not include any
luminosity error. In Ref.\cite{JLC1} it has been argued that the
luminosity can be measured with an error of ${\cal O}(1\%)$. On the
other hand, using a typical $\se$ production cross section $0.1$
pb$\sim 0.2$ pb (See.Fig 2), and assuming a constant acceptance of
27\%, we can see that the errors on  $M_1$ are estimated based on
${\cal O}(1000)$ accepted $\se$ events for the luminosity of 20
fb$^{-1}$ (Fig.11), and ${\cal O}(5000)$ events for 100 fb$^{-1}$
(Fig.12). This corresponds to an error on the cross section of about
3\% and 1.4\%, respectively. The latter is already comparable to the
luminosity error. Hence a further increase of statistics would not
improve the actual error on the MSSM parameters unless the error on
the luminosity is also reduced further.

The estimated errors on  $\tan\beta$ are rather impressive, compared to
those from the other experimental methods. In Fig.13, we plot the
error on $\tan\beta$ which can be obtained from the lighter chargino
($\tilde{\chi}^+_1$) production and the co-production of
$\tilde\chi^\pm_1$ and $\tilde\chi^\mp_2$. Here use has been made of
the direction of pair produced charginos $\tilde{\chi}^+_1$ with a
10\% acceptance from its decay products, assuming $\int{\cal L}dt= 100$
fb$^{-1}$ for both $P_e=1$ and $0$.\footnote{The 
direction of a produced chargino can be solved for with
a two-fold ambiguity when the chargino decays into
$W\tilde{\chi}^0_1$\cite{TSUKA}.
The forward backward asymmetry for the final state $W$ can also be
used even if the $\tilde{\chi}^+_1$ decays into $W^*
\tilde{\chi}^0_1$\cite{FLEP,FLC}.  } The errors on
$m_{\tilde{\chi}^\pm_1}$ and $m_{\tilde{\chi}^\pm_2}$ were 
both assumed to be
2\% as long as they are accessible kinematically.\footnote{
$\Delta{m_{\tilde{\chi}^\pm_1}}$ was found to be around 5\% for 50
fb$^{-1}$ of data\protect\cite{TSUKA}. A threshold scan for the
$\tilde{\chi}^+_1\tilde{\chi}^-_1$ pair production might determine the
mass better.} One can see that the upper bound practically disappears
when $\tan\beta$ exceeds 5.\footnote{It has been claimed that a very
precise measurement of $\tan\beta$ is possible when $\tan\beta\sim
4$\protect\cite{FLC} if the chargino mass errors are
negligibly small. Some additional error on $\tan\beta$ has been
introduced here assuming a finite error on $m_{\chi^{\pm}_i}$.
In Fig.13, we have also taken larger value of $m_{\tilde\nu}$
compared to \protect\cite{FLC}, where sensitivity to 
$\tan\beta$ is smaller.}

\subsection*{4.3 Checking Supersymmetry Relation}

So far we have been assuming that new particles found at an LC are
superpartners of leptons. In other words, we have implicitly been
using supersymmetric interactions of sfermions with the neutralinos of
the MSSM without any attempt at checking the nature of the
interactions. Instead, we merely used the data to determine the free
parameters of the MSSM, such as $M_1$ and $\tan\beta$. In this
subsection we are going to discuss the possibility to probe the
gaugino-sfermion-fermion interaction (more specifically, the
$\tilde{B}$-$\tilde{e}_R$ -$e_R$ coupling), and some aspects of the
$\tilde{B}$-$\sti$-$\tau$ coupling.

We start our discussion with $\tilde{e}_R$. The production proceeds
though the $s$-channel exchange of gauge bosons and $t$-channel
exchange of neutralinos, whose cross section is shown in Appendix
B. The $t$-channel exchange is dominated by bino-like neutralino
exchange, which led us to the simple dependence of the cross section
on the gaugino mass $M_1$ as has been shown in Fig.3a).

The tree level coupling of the $\tilde{B}$-$\tilde{e}_R$-$e_R$ vertex
has a simple relation to the $B$-$e$-$e$ coupling in the MSSM:
\be\label{e18} 
g_{\tilde{B}\tilde{e}_R e_R } =\sqrt{2} g\tan\theta_W=\sqrt{2}g'. 
\ee 
This relation is imposed by supersymmetry. Thus the measurement of
$g_{\tilde{B}\tilde{e}e }$ will allow us to prove that $\tilde{e}$ and
$\tilde{B}$ are indeed superpartners of $e_R$ and $B$.

 For this test we modify the relation of Eq.(\ref{e18}) as
\be\label{e19} g_{\tilde{B}\tilde{e}_R e_R}=\sqrt{2}g' Y_{\tilde B}.
\ee and estimate the sensitivity to $Y_{\tilde B}$ by introducing a
new $\dbarchi$ function for the selectron pair production which
depends on $Y_{\tilde B }$ though $g_{\tilde{B}\se_R e}$.  In the
limit of $m_Z\ll M_1$ and $\mu$, we obtain an approximate formula for
the matrix element ${\cal M}$: \be\label{e20} {\cal M}\propto
\sin\theta\left[
1-\frac{4Y_{\tilde{B}}^2}{1-2\cos\theta\beta_f+\beta_f^2+4M_1^2/s}
\right] .  \ee It is apparent from Eq.(\ref{e20}) that one can
constrain both $Y_{\tilde B}$ and $M_1$ by measuring the differential
cross section: $d\sigma(e^+e^-\rightarrow\se^+_R\se^-_R)/d\cos\theta$.

Figure 14 is a $\dbarchi_{\se}$ contour plot projected on the
$M_1$-$Y_{\tilde{B}}$ plane for a representative point in the
parameter space of the MSSM: $m_{\se_R}=200$ GeV, $\mu=300$ GeV,
$M_1=99.57$ GeV and  $\tan\beta=2$.  One finds a good sensitivity to the
coupling $Y_{\tilde{B}} $ of ${\cal O}(1\%)$ in this case. The reason
why we got upper and lower bounds on $M_1$ and $Y_{\tilde{B}}$ is as
follows: when we increase $M'_1$ from $M_1$ to $M_1+\Delta M_1$, the
total cross section decreases. The corresponding increase of
$\dbarchi_{\se}$ can be compensated by increasing $Y_B$
correspondingly. However, due to the constraint on $\mchi$ which comes
from the electron energy distribution from decaying $\tilde{e}_R$'s,
the optimized value of $\mu'$ is smaller for a larger $M_1'$ (see
Fig.2 for the relation between $M_1$ and $\mu$). Hence the second
lightest neutralino mass is lighter for a larger $M_1'$, and it also
tends to have a larger mixing with $\tilde{B}$. At the $M'_1$ upper
bound of $\dbarchi=1$ , the polar angle distribution changes its shape
so that it is less forwardly peaked. On the other hand, the lower
bound on $M'_1$ is determined by $\mchi$. Namely, for given $M_1$,
there is a upper bound on $\mchi$ which one can obtain by varying
$\mu$ and $\tan\beta$ maximally. It becomes thus harder to reproduce
the input $\mchi$ as we decrease $M'_1.$

Some deviation of $Y_{\tilde{B}}$ from its tree level value is
expected if we take into account the effect of radiative corrections in
the framework of the MSSM. If there is a large difference between
several soft SUSY breaking mass parameters, such a correction occurs
very naturally. For example, if $m_{\tilde q}\gg m_{\tilde l}$ and
$m_{\tilde{\chi}}$, the effective theory below $Q<m_{\tilde{q}}$ is
not supersymmetric, and couplings related by supersymmetry start to
run differently according to the RG equations of the effective theory.
In particular, both squarks and quarks decouple from the wave function
renormalizations of gauginos in the low energy effective theory, while
only squarks decouple from that of gauge bosons, from which
$Y_{\tilde{B}}\neq 1$ may originate.

The RG equations below the squark decoupling are as
follows\cite{CHAN}: \ben\label{e21}\beq \frac{d\alpha'}{d\log
Q}&=&\frac{55}{12\pi}\alpha'^2, \\ \frac{d \alpha_{\tilde B \tilde{e}e }}
{d\log Q} &=&\frac{11}{4\pi}\alpha'\alpha_{\tilde B \tilde{e}e },
\eeq\een where we neglected terms proportional to $(Y_{\tilde B}-1)$
on the right-hand side of the equations. We find from Eq.(\ref{e21})
\be\label{e22}\Delta Y_{\tilde B e \tilde{e}} /Y_{\tilde B e \tilde{e}}
=0.007t_{\tilde{q}\tilde{e}}, \ee where
$t_{\tilde{q}\tilde{l}}=\log_{10}(m_{\tilde{q}}/m_{\tilde{l}})$.

It is rather striking that the error on the couping is of about the
same order as that of the radiative correction proportional to
$\log(m_{\tilde{q}}/m_{\tilde{l}})$ if the squark mass is much heavier
than the slepton mass.  This, on one hand,
requires knowledge of $m_{\tilde{q}}$ and a full 1-loop calculation
of the process to remove the uncertainty in  $Y_{\tilde B}$ from the
determination of $M_1$; notice that the error on $M_1$ increases by a
factor of two, if we let $Y_B$ move freely. This also implies a larger
error on $\tan\beta$, as the errors on $M_1$ and $\tan\beta$ are
correlated strongly when the lightest neutralino is gaugino-dominant.
On the other hand, we can turn this argument around. Then emerges the
possibility to constrain the squark mass scale from the measurement of
$Y_{\tilde B}$ or other couplings even if the energy of future
colliders is not enough for the squark production. A full calculation
of 1-loop radiative corrections to this process is eagerly anticipated.

A similar radiative correction to $g_{\tilde{W}\tilde{\nu} e_L }$
turns out to be of the order of 2\%$\times
\log_{10}(m_{\tilde{q}}/m_{\tilde{l}})$, but the sensitivity to this
coupling has been argued to be rather poor\cite{FLC}: about $-$
15\%+30\% for a representative parameter choice. This estimate is
based on the study of gaugino-dominant chargino production and decay,
using the forward-backward asymmetry of the decay products and the
total production cross section. The chargino production proceeds through
$t$-channel exchange of $\tilde{\nu}$, and the $\tilde{\nu}$ is
assumed kinematically inaccessible in the study. Its mass is
determined by comparing the production cross sections for  polarized
and  unpolarized electron beams, and the decay forward-backward
asymmetry, but it has a very large uncertainty. Furthermore, the
uncertainty in  the branching ratios introduces  a systematic error to the
measurement of the total cross section. Notice that the production
cross section of gaugino-dominant charginos is very small for a
right-handed polarized beam, hence the chargino study heavily relies
on the use of the  unpolarized beam, where large backgrounds limit 
the decay modes to study. The estimated error on 
the cross section is about 5\%
for $\int{\cal L}dt =100$ fb$^{-1}$. These uncertainties limited the
$g_{\tilde{W}\tilde{\nu}e}$ study.\footnote{ During this work, we
learned of similar work on measuring decoupling effects by
H.-C. Cheng, J.L. Feng, and N. Polonsky\protect\cite{CFP}. We thank Jonathan
Feng for bring this work to our attention.}

The coupling $g_{\tilde{B}\tilde{e}_Re}$ is considerably easier to
measure than $g_{\tilde{W}\tilde{\nu}e}$. $\se_R$ has a sizable
production cross section for the  right-handed electron  beam, and the
uncertainty in  the production cross section is expected to be very
small if the $\se$ decays exclusively into $e\tilde{\chi}^0_1$. For the
representative parameters of Fig.14, the $t$-channel exchange of
neutralinos is dominated by the lightest one, $\tilde{\chi}^0_1$ , and
its mass is well constrained from the $\se_R$ decay data. It is
therefore not surprizing that the coupling is  measured very well here.

Deviations from tree level MSSM predictions can also appear in other
couplings involving sleptons or neutralinos, such as
$g_{\tilde{H}l\tilde l}$. Unfortunately, the measurement of
$\tan\beta$ has a large error, thus radiative corrections may not be
relevant in this case. The gaugino mass matrix also gets radiative
corrections to its tree level value\cite{many}. If one assumes a
unified gaugino mass at the GUT scale, one may in principle extract
the squark mass scale from the gaugino mass relation at the weak
scale. Unfortunately, the measurements of gaugino masses are limited
by ambiguities in the neutralino and chargino mixing angles
\cite{JLC1,TSUKA,FLC}. The gluino mass, though involving no mixing, 
is hard to measure precisely at hadron colliders, too.\cite{GLUINO}

Now we turn our attention to $\sti$ production. In the previous
subsection, we found that
$P_{\tau}(\sti\rightarrow\tau\tilde{\chi}^0_1)$ becomes independent of
$\tan\beta$, if $\tilde{\chi}^0_1$ is gaugino-dominant, or in other
words $m_{\tilde{\chi}^0_1}\sim M_1$. In such a situation,
simultaneous measurements of $\theta_{\stau}$ (using the total
$\sti$-pair production cross section) and $P_{\tau}$ constrain
the nature of the $\tilde{B}$-$\tilde{\tau}$-$\tau$ coupling instead of
constraining $\tan\beta$. Given the fact that $\tilde{\chi}^0_1$ is
almost a pure gaugino (which can be checked with scalar electron
production), the measurement of the total $\sti$ pair production cross
section essentially fixes the polarization $P_{\tau}$ through
Eq.(\ref{e8}). Any deviation of $P_{\tau}$ from it indicates that
something unexpected is happening.

In Fig. 15 we show $\dbarchi=1$ contours by taking the mixing angle
parameter ($\theta_{\stau}$) in the $\tilde{\chi}\tilde{\tau}\tau$
coupling free from that in the $Z\tilde{\tau}\tilde{\tau}$ coupling
($\bar{\theta}_{\stau}$). We can say that
$\theta_{\stau}-\bar\theta_{\stau}$ measures the chirality flipping
part of the $\tilde{B}(\tilde{W})$-$\tilde{\tau}$-$\tau$ interaction
which is zero in the MSSM. Due to the dependence of $P_{\tau}$ on
$\tan\beta$ through a small but finite higgsino component in the
neutralino mass eigenstate ($\tilde{\chi}^0_1$), the sensitivity of
the $\tilde{\tau}\tau\tilde{\chi}^0_1$ coupling to the
$\theta_{\stau}$ is marginal unless $\mu\gg M_1$.\footnote{ The
contours in Fig.15 depend sensitively on the region of $\tan\beta$
searched. We took $1<\tan\beta<50$ here to obtain $\dbarchi=1$,
assuming Yukawa coupling is not too large at GUT scale.}

Nevertheless the figure can be regarded as an example of a ``no lose
theorem'' of precision measurements of supersymmetry processes.
Depending on the position in the parameter space of the model, we
occasionally lose sensitivity to some parameter, as we have seen for
the $\tan\beta$ determination using slepton production. We can,
however, turn this into an advantage: the process becomes independent
of the ambiguity caused by the parameter, and we can test its
supersymmetric nature. In the current case, we can check the chiral
nature of the gaugino, thanks to the insensitivity of the process to
$\tan\beta$.

\section*{5. Conclusion}
In this paper we presented an extensive study of the production and
decay of the lighter scalar tau lepton $\sti$ at a future LC and
discussed physics that could be extracted from them. Studying $\stau$
production
is important because it may be lighter than the other sleptons,
and could thus be found earlier. The light $\sti$ case is also
theoretically well motivated in the MSUGRA-GUT model, and is not
excluded, at least, in other models, as long as there is a large
$\tilde{\tau}_L$-$\tilde{\tau}_R$ mixing.

We discussed that the $\stau$ mass matrix at $\mgut$ might provide a
clue to distinguishing SUGRA-GUT from DNNS models. In order to obtain
the GUT scale mass matrix, one must know the mass matrix at the weak
scale and the tau Yukawa coupling ($Y_{\tau}$) which is characterized
by $\tan\beta$.

The mass matrix can be determined if one knows the $\stau$ masses and
mixing angle $\theta_{\stau}$. The feasibility of determining those
parameters at an LC has been studied for the lighter mass eigenstate $\sti$,
assuming the JLC1 model detector.
For a representative parameter set: $\msti=150$ GeV and $\mchi =100$
GeV, we found that these masses can be measured to $\Delta\msti=4.1$
GeV and $\Delta\mchi=3$ GeV for $10^4$ $\sti$ pairs produced with
background corresponding to $\int {\cal L}dt =100$ fb$^{-1}$, assuming
that $\sti$'s decay exclusively into $\tilde{\chi}^0_1\tau$. For the
same mass parameters and luminosity conditions, the expected
statistical error on the mixing angle turned out to be
$\Delta\sin\theta_{\stau}=0.045$, when $\sin\theta_{\stau}=0.75$.

The polarization ($P_{\tau}$) of $\tau$ leptons from $\sti$ decay is
sensitive to $\tan\beta$ because of its dependence on the $\tau$ Yukawa
coupling. The expected statistical error on the polarization was
estimated to be $\pm 0.07$ for $10^4$ $\sti$ pairs and background
corresponding to $\int {\cal L}dt =100$ fb$^{-1}$. Using the information
of the neutralino mass matrix obtained from the simultaneous studies
of the $\tilde{e}_R$ production and decay, $\tan\beta$ might be
determined. The error on $\tan\beta$ varies drastically with $M_1$ and
$\mchi$, as shown in Fig.12 for some representative points in the
parameter space of the MSSM.

Notice that $\tan\beta$ is one of the most important parameters that
determine the Higgs sector of the MSSM. At the same time, it is known
to be difficult to measure especially if it is large. If
$\tan\beta>10$, $\sti$ decays give us a unique opportunity to
determine $\tan\beta$.

We have also discussed a possibility to test the supersymmetry
relations among couplings involving superpartners.  By studying the
polar angle distribution of $\se_R$ production, one can measure not
only $M_1$, $\mchi$ but also the gaugino-selectron-electron coupling
$g_{\tilde{B}\tilde{e}_Re}$. A fit allowing $g_{\tilde{B}\tilde{e}_R
e}$ to move freely from the tree level prediction of supersymmetry
gives $\Delta g_{\tilde{B}\tilde{e}_R e}$ $\sim {\cal O} (1\%\sim
2\%)$. This is comparable to typical radiative corrections to the same
coupling $\sim 0.7\%$ $\times\log(m_{\tilde{q}}/m_{\tilde{l}})$. This
suggests that the LC might allow us to start
probing radiative corrections to couplings involving SUSY particles.

Implications of the MSUGRA model at LEPII and LHC have been discussed
and studied in many papers. Unfortunately, prospects to
determine the soft SUSY breaking mass parameters are not so bright
there: as for LEPII, its available luminosity is too low for the
slepton study and one has to fight the enormous background coming from
$W^+W^-$ production for chargino study\cite{FLEP}, while at LHC, one
suffers from the high QCD background even though strongly interacting
superparticles will be copiously produced. Therefore those studies in
the framework of the MSUGRA model are focused mostly on the discovery
potential of the machine in question.  However, it is becoming more
and more recognized that we can certainly go beyond that if a next generation
linear $e^+e^-$ collider is actually built. Namely, the
experiments at the LC will make it possible to measure the parameters
of the MSSM once a superparticle is discovered, which will then enable
us to check the predictions of the models of SUSY breaking.

\section*{Acknowledgement}
One of the author (M.M.N) have been benefitted from  an unfinished 
collaboration with J. Hisano and Y. Yamada. The 
authors  are greatful to  M. Tanaka for 
checking $e^+e^-\tau^+\tau^-$ background  
cross section independently. The authors also  
thank M. Drees and K.Hikasa on reading 
manuscript carefully. This work is supported 
in part by the Grant in aid for 
Science and Culture of Japan(07640428). 

\section*{Appendix A. Parameters of the MSSM}

In this subsection, we are going to summarize the interactions of
$\stau$ that are relevant for the analysis in this paper. The
interactions are fixed by supersymmetry and gauge symmetry, as well as
by the mass parameters of the model.
 
The $\stau$-$\tau$-ino interactions relevant to $\stau_i$ decay are
expressed by the following Lagrangian \cite{NOJIRI}:
\newenvironment{sarray}{\renewcommand{\arraystretch}{0.5}
\begin{array}}
{\end{array} \renewcommand{\arraystretch}{1}}
\be\label{a1}
{\cal L}=
\sum_{i=1,2\  j=1,..,4}
\tilde{\tau}_i\  \bar{\tau}( P_L a^R_{ij}+P_R a^L_{ij})\tilde{\chi}^0_j
+ \sum_{i=1,2\  j=1,2}
\tilde{\tau}_i\  \bar{\nu}_{\tau} P_R b_{ij}\tilde{\chi}^+_j\ \  +\ \  h.c.,
\ee
where
\ben\label{a2}\beq
&\left( \begin{array}{c}a^{R(L)}_{1j}\\ a^{R(L)}_{2j}\end{array}\right)
&=
\left(\begin{array}{cc}\cost&\sint\\
-\sint&\cost \end{array}\right)
\left(\begin{array}{c}a^{R(L)}_{Lj}\\ a^{R(L)}_{Rj}\end{array}\right),
\nonumber
\\
&\left( \begin{array}{c}b_{1j}\\ b_{2j}\end{array}\right)
&=
\left(\begin{array}{cc}\cost&\sint\\
-\sint&\cost \end{array}\right)
\left(\begin{array}{c}b_{Lj}\\ b_{Rj}\end{array}\right),
\eeq
\beq
&a^R_{Lj}=-\frac{g m_{\tau}}{\sqrt{2}m_W \cos\beta}N_{j3},
\ \ 
&a^L_{Lj}=\frac{g}{\sqrt{2}}\left[ N_{j2}+N_{j1}\tan\theta_W\right],
\nonumber
\\
&a^R_{Rj}=-\frac{2g}{\sqrt{2}}N_{j1}\tan\theta_W,
\ \ 
&a^L_{Rj}=-\frac{g m_{\tau}}{\sqrt{2}m_W \cos\beta}N_{j3},
\nonumber\\
&b_{Lj}=-gU_{j1},
\ \ 
&b_{Rj}=\frac{gm_{\tau}}{\sqrt{2}m_W \cos\beta } U_{j2}.
\eeq
\een

Here the real orthogonal matrix $N_{ij}$  and 
unitary matrices $U_{ij}$ and  $V_{ij}$ are  the 
diagonalization matrices of  the neutralino mass matrix ${\cal M}_N$ 
and chargino mass matrix ${\cal M}_C$ as follows:
\be\label{a3}
U^*{\cal M}_C V^{-1}=M_C^D, \ \ N {\cal M}_N N^{-1}=M^D_N, 
\label{a3a}
\ee
where the mass matrices are written in the following form:
\ben\label{a9}
\beq
&&{\cal M}_N(\tilde{B},\tilde{W_3},\tilde{H_1},\tilde{H_2})=
\nonumber\\
&&\left(\begin{array}{cccc}
M_1 &0&-m_Z\sw\cosb&m_Z\sw\sinb\\
0 &M_2& m_Z\cw\cosb&-m_Z\cw\sinb\\
-m_Z\sw\cosb&m_Z\cw\cosb&0&-\mu \\
m_Z\sw\sinb&-m_Z\cw\sinb&-\mu & 0
\end{array}\right),\nonumber\\
\label{a9a}
\\
&&{\cal M}_C(\tilde{W},\tilde{H})=\left(\begin{array}{cc}
M_2 & m_W \sqrt{2}\sinb\\
m_W\sqrt{2}\cosb & \mu
\end{array}
\right).
\label{a9b}
\eeq
\een
 $\mu$ is a supersymmetric higgsino mass parameter, while $M_1$ and
$M_2$ are the soft breaking mass parameters of Bino and Wino
introduced previously.  The mass eigenstate $\tilde{\chi}^0_i$ and
current eigenstates $\tilde{B},\tilde{W},\tilde{H}_1,\tilde{H}_2$ are
related by
\be
\tilde{\chi}^0_i=N_{i1}\tilde{B}+N_{i2}\tilde{W}+N_{i3}\tilde{H}_1
+N_{i4}\tilde{H}_2.
\ee 

Unlike the notation of Haber and Kane\cite{SUSY}, we take $N$ to be
real so that $m_{\tilde{\chi}_i^0}$ can be either positive or
negative. Its sign must be kept to understand the equations in
Ref{\cite{NOJIRI}}.  We take $\vert m_{\tilde{\chi}_1^0} \vert \leq $
$\vert m_{\tilde{\chi}_2^0} \vert \leq $ $\vert m_{\tilde{\chi}_3^0}
\vert \leq $ $\vert m_{\tilde{\chi}_4^0} \vert$ and $0\leq
m_{\tilde{\chi}_1^-} \leq $ $m_{\tilde{\chi}_2^-}$. We assume the mass
relation of MSUGRA $M_1=(5/3)\cdot \tws M_2$ for numerical
calculations in order to reduce the number of parameters.

 Eq.(\ref{a1}) leads to an expression for  $P_{\tau}$:
\be P_{\tau}(\sti\rightarrow\tilde{\chi}^0_1\tau)=
\frac{(a_{11}^R)^2-(a_{11}^L)^2}{(a_{11}^R)^2+(a_{11}^L)^2}
\ee

\section*{Appendix B $\se_R$ production}
$\se_R$ production proceeds through $t$-channel exchange 
of neutralinos and $s$-channel exchange of gauge bosons.
The tree level couplings of 
the $\tilde{e}_R e \tilde{\chi}_i^0$ vertices 
may be read off  from 
Eqs.(\ref{a1}) and (\ref{a2}) by setting $\sin\theta_{\stau}=1$ and 
 replacing $\tau\rightarrow e$. We obtain the formula for 
the $\se_R$-pair production cross section  as follows:
\ben\label{b1}
\beq
\frac{d\sigma}{d\cos\theta}(h_e, \bar{h}_e)
&=&\frac{1}{2s}
\frac{\beta_f}{16\pi}\frac{1}{2}\cdot \sum_{{\bar h}_e}\vert{\cal M}
(h_e, \bar{h}_e)\vert^2\\
i{\cal M}(h_e, \bar{h}_e)&=&-i\lambda_ie^{i\lambda_i\phi}\sin\theta 
s\beta_f \left[g_Z^2 \frac{A_{h_e}A{\frac{1}{2}}}
{s-m_Z^2+i\Gamma_Z}\right.\nonumber\\
&&\left.+\frac{e^2}{s}+\frac{(1\pm (-)^{\bar{h}+\frac{1}{2}})}{2}
\sum_{j}{1 \over 2}\frac{A_{j R}^2}{t-m_{\tilde{\chi}_j}^2}\right],
\eeq\een
where $h_e(\bar{h}_e)=$ $\pm 1/2$ 
represents the helicity of the initial-state electron (positron),
$\lambda_i\equiv h_e-\bar{h}_{e}$, $\theta$ and $\beta_f$ 
is the $\se^-_R$ production angle and velocity, 
and $t=-\frac{s}{4}(1-2\cos\theta\beta_f+\beta_f^2)$. 
The couplings $A_{h_e}$ and $A_{jR(L)}$ are given by  
\beq
&A_{1/2}= \sws, \ \ \ \ &A_{-1/2}=-\frac{1}{2}+\sws \nonumber\\ 
&A_{jR}=-\sqrt{2}g\tan\theta_W N_{j1}. 
\eeq

%
%
\begin{figure}
\caption{
Interactions of neutral components of gauginos and higgsinos with $\stau_R$
and $\tau_L$ or $\tau_R$.} 
\end{figure}
\begin{figure}
\caption{
$\mchi=100$ GeV contours in the $M_1$-$\mu$ plane: solid and dotted
lines correspond to $\tan\beta=1.5$ and 30, respectively.}
\end{figure}
\begin{figure}
\caption{
a):$\sigma_{\tilde{e}_R^+\tilde{e}_R^-}$ contours with 
$m_{\tilde{e}}=200$ GeV, $\protect\sqrt{s}=500$ GeV, and 
 $P_e=1$ in the $M_1$-$\tan\beta$ plane. 
At each point of the figure, $\mu$ is chosen so that 
$\mchi=100$ GeV. Solid lines correspond to a  $\mu>0$ 
solution and the dashed lines to $\mu<0$. b):
 $P_{\tau}(\stau_R\rightarrow\tau\chi^0_1)$ 
contours 
in the $M_1$-$\tan\beta$ plane
with the same neutralino mass constraint. Only the contours of 
 $\mu>0$ solutions are shown. }
\end{figure}
\begin{figure}
\caption{ 
a) and b): Energy distributions of the $\rho$ and $\pi$ from 
a) $\tau_R$ decay and b) $\tau_L$  decays with a fixed $E_{\tau}$
($E_{\tau}\gg\tau$)
as  functions of $z\equiv E_{\pi(\rho)}/E_{\tau}$.
c) and d):  Energy distributions of the $\rho$ and $\pi$  from 
a cascade decay of a $\stau$ for $\msti=150$ GeV, $\mchi=100$ GeV and 
$\protect\sqrt{s}=500$ GeV. The  
$\stau$ decays exclusively into $\tau^-_{R(L)}$ in 
c)(d)).
}
\end{figure}
\begin{figure}
\caption{ 
$\sla{P}_T$ distributions of events passing cuts 1)-5), for
$\protect\sqrt{s}=500$ GeV, $\msti=150$ GeV, and $\mchi=100$ GeV. The
solid and dotted lines with higher $\sla{P}_T$ tails are for $10^4$ $\stau$
pairs decaying exclusively into $\tau^-_L$ and $\tau^-_R$,
respectively. The $\sla{P}_T$ distribution of the $ee\tau\tau$ background,
also shown in the figure, corresponds to $\int{\cal L}dt =100$
fb$^{-1}$.  }
\end{figure}
\begin{figure}
\caption{
$E_{\rm jet}$ distributions of $10^5$ $\sti$ pairs decaying
exclusively to $\tau_L$ for different total $\sla{P}_T$ cuts: no 
$\sla{P}_T$ cut
(dotted), $\sla{P}_T>15$ GeV (solid), $\sla{P}_T>35$ GeV(dashed).  
$\sla{P}_T>35$ GeV is
the optimal cut for $\theta^{veto}_e =150$ mrad, while $\sla{P}_T>15$ GeV
for $\theta^{veto}=50$ mrad.  }
\end{figure}
\begin{figure}
\caption{ 
a):Invariant mass distributions of jets consisting of $\pi^-$ + one or
two $\gamma$'s for $\stau^+\stau^-$ $\rightarrow \tau^+\tau^-
\tilde{\chi}^0_1\tilde{\chi}^0_1$, followed by  
$\tau^+\tau^-\rightarrow \pi^+\rho^-\nu_{\tau}\bar{\nu_{\tau}}$
(solid) and $\pi^+a_1^-\nu_{\tau}\bar{\nu}_{\tau}$(bars).  The latter
corresponds to misidentified $a_1^-\rightarrow
\pi^-\pi^0\pi^0\rightarrow \pi^- 4\gamma$. We assumed the parameters
of the JLC1 model detector. The invariant mass distribution for the
jets from $a_1$ 3-prong decays ($a_1^-\rightarrow 2\pi^-\pi^+$) is
also shown as the dotted line. Due to the photon mis-measurements the
jet invariant mass distribution of $a_1$ 1-prong decays sits below the
one for 3 prong decays.  b) The jet energy and c) $z_c$ distributions
for the events that satisfy $\rho$ cuts. The solid line is the
distribution for $\stau^+\stau^-\rightarrow \pi^+\rho^-$, while bars
are of $\stau^+\stau^-\rightarrow\pi^+a^-_1$ as before.  The
contamination is larger for small $z_c$ or higher $E_{\rm jet}$. }
\end{figure}
\begin{figure}
\caption{ 
Results from the mass fit to $10^4$ $\sti\sti$ pair events decaying
into $\tau_R\tilde{\chi}_1^0$ exclusively, where the SM background
corresponding to $\int{\cal L}dt=100$ fb$^{-1}$ has been included in
the fit: a): the jet energy distribution for the $\rho$ events
selected from data MC events (bars) and the best fit histogram. In the
fit we kept $P_{\tau}=+1$ and normalized the histogram so that the
total number of events agreed with that of the MC data.  The average
SM background is also shown in the figure.  b) Contours for
$\Delta\chi^2=1$ and 4 in the $\mchi$ and $\msti$ plane.  }
\end{figure}
\begin{figure}
\caption{ 
$z_c$ distributions for the $\rho$ candidates selected from $10^4$
$\sti$ pairs decaying exclusively into a):$\tau_R$ and b):$\tau_L$ ,
together with the best fit histogram. The sample with $E_{\rm jet}>20$
GeV and $0.08\le z_c\le 0.92$ are used for the fit.  The best fit
values of $P_{\tau}$ and their errors were obtained to be
$0.995\pm0.082$ and $-0.991\pm0.008$ for $\tau_R$ and $\tau_L$,
respectively, for fixed $\msti=150$ GeV and $\mchi=100$ GeV.}
\end{figure}
\begin{figure}
\caption{ 
$\Delta\chi^2$ contours in the $\msti$-$\sin\theta_{\stau}$ plane,
resulting from the fit to 5000 $\sti$ pair generated for $\int {\cal
L}dt =100$ fb$^{-1}$, $\msti=150$ GeV, $\mchi=100$ GeV, and
$P_{\tau}=0.6788$. The MC sample corresponds to a $\sti$ with
$\sin\theta_{\tau}=0.7526$ decaying exclusively into a bino-like
lightest neutralino.  }
\end{figure}
\begin{figure}
\caption{ 
1-$\sigma$ error band on $M_1$ estimated using the $\dbarchi_e$
function. Solid and dotted lines plot values of $M'_1-M_1$ in \%,
where $\dbarchi_{\se}\vert_{\rm min}=1$ for the $M'_1$ varying other
fitting parameters ($m'_{\se}$, $\mu'$, $\tan\beta'$).  Input values
are chosen so that $\mchi=100$ GeV and $\mu$ is positive for each
$M_1$. Solid lines are for $\tan\beta=1.5$ as input where the
parameter region with $1<\tan\beta'<100$, and $50$ GeV$<\mu'<10^4$GeV
was searched to obtain $\Delta M_1$. Dashed lines for $\tan\beta=15$,
but only the parameter region with $\tan\beta>10$ are searched to
obtain these 1-$\sigma$ limits on $M_1$.  }
\end{figure}
\begin{figure}
\caption{ 
$\Delta\bar{\chi}^2=\Delta\bar{\chi}^2_{\se}$
$+\Delta\bar{\chi}^2_{\stau}=1$ contours projected onto the
$M_1$-$\tan\beta$ plane.  Projections of the contours on $M_1$ or
$\tan\beta$ corresponds to 1-$\sigma$ errors of the parameter.  Input
values are chosen to be $\msti=150$ GeV, $\sin\theta_{\stau}=1$
($\sti=\stau_R$), $\mchi=100$ GeV( $\mu>0$), and $m_{\tilde{e}_R}=200$
GeV. Input values of $\mu$ are explicitly shown in the figure for
individual sample points.  $P_{\tau}(\stau_R\rightarrow\tau
\tilde{\chi}^0_1)=0.8, 0.4, 0,$ and $-0.4$ contours with $\mchi=100$
GeV and $\mu>0$ are also plotted in the figure as dashed lines.  }
\end{figure}
\begin{figure}
\caption{ 
1(2)-$`\sigma$' errors on $\tan\beta$ from chargino production as
functions of input $\tan\beta$.  We used chargino distributions for
$P_e=+1$ and $0$ with $\int{\cal L}dt =100$ fb$^{-1}$, and assuming
that both chargino masses are known to 2\%  accuracy.  Upper bound
practically disappears when $\tan\beta$ exceeds 5. Input values 
are $M_2=210$GeV, $\mu=-195$ GeV and $m_{\tilde{\nu}_2}=500$ GeV.}
\end{figure}
\begin{figure}
\caption{ 
$\Delta\bar{\chi}^2_{\se}=1$ contour in the $M_1$-$Y_{\tilde B}(\equiv
g_{\tilde{B}\tilde{e}_R e}/$ $\protect\sqrt{2}$ $g')$ plane.  The
definition of $\Delta\bar{\chi}^2_{\se}$ has been modified to allow
$g_{\tilde{B}\tilde{e}_Re}$ to deviate from $\protect\sqrt{2}$
$g'$. Input values are $m_{\tilde{e}_R}=200$ GeV, $\mu=300$ GeV,
$M_1$=99.57 GeV, and $\tan\beta=2$. The error on the coupling is of
about the same order as that of the radiative correction proportional
to $\log(m_{\tilde{q}}/m_{\tilde{l}})$ when $m_{\tilde{q}}
/m_{\tilde{l}}\sim 10$.  }
\end{figure}
\begin{figure}
\caption{
$\Delta\bar{\chi}^2=1$ contours when the mixing angle $\theta_{\stau}$
in the $\tilde{\chi}^0_1\tilde{\tau}\tau$ is allowed to move freely
from that in the $Z\tilde{\tau}\tau$ coupling
($\bar\theta_{\stau}$). $\theta_{\stau}-\bar\theta_{\stau}$
parametrizes the chirality flipping part of $\tilde{B}(\tilde{W})$-
$\stau$-$\tau$ interaction, which is zero in the MSSM.  The solid line
corresponds to $\mu=-600$ GeV and the dashed line to $\mu=-200$
GeV. The other parameters are fixed to $M_1=104.5$ GeV, and
$\tan\beta=20$.  The sensitivity is moderate for $\mu\gg M_1$.}
\end{figure}
\begin{table}
\begin{center}
\begin{tabular}{|c||c|c|c|}
process &$\sigma_{P_e=+0.95}$(fb) & $\sigma_{\rm cut}$ (fb) 
& number of   $\rho$ candidates/(100 fb$^{-1}$) \\ 
\hline\hline
$WW\rightarrow \tau^+\tau^-$&6.23 & 0.16& 14.9\\
\hline
$eeWW\rightarrow \tau^+\tau^-$& 2.16 & 0.21& 17.9\\
\hline
$ZZ\rightarrow \tau^+\tau^-\nu\bar{\nu}$ & 4.88 &0.59& 51.0\\
\hline
$\nu\bar\nu Z \rightarrow\tau^+\tau^-$ & 0.46 & 0.07& 6.3\\ 
\end{tabular}
\end{center}
\caption{
Dominant background cross sections at $P_{e}=+0.95$ to 
the process $e^+e^-\rightarrow\sti\sti$ 
followed by  $\sti\rightarrow\tilde{\chi}^0_1\tau$. 
Background cross sections  after requiring the cuts described in the text, 
and the average number of $\rho$ background events for 
$\int{\cal L}dt=100$fb$^{-1}$ are also shown in the table.}
\end{table}

\end{document}